\documentclass[11pt,superscriptaddress,aps,prd]{revtex4}

\everymath{\displaystyle}
\usepackage{graphicx,amsmath,amssymb,amsfonts,amstext,latexsym,stmaryrd,amscd,txfonts,pxfonts,hyperref}
\usepackage{lineno,hyperref}
\usepackage{siunitx}
\usepackage{cmap}				% Mapear caracteres especiais no PDF
\usepackage{lmodern}			% Usa a fonte Latin Modern
\usepackage[T1]{fontenc}		% Selecao de codigos de fonte.
\usepackage[utf8]{inputenc}		% Codificacao do documento (conversão automática dos acentos)
\usepackage{indentfirst}		% Indenta o primeiro parágrafo de cada seção.
\usepackage{color}				% Controle das cores
\usepackage{graphicx}			% Inclusão de gráficos
\usepackage{amsmath}
\usepackage{mathtools}
\usepackage[english]{babel}
\usepackage{appendix}
\usepackage{verbatim}
\usepackage{float}
\usepackage{subfigure}
\usepackage{soul}
\hypersetup{
    colorlinks=true,
    linkcolor=blue,
    filecolor=magenta,      
    urlcolor=black,
		citecolor=red,
		}

\begin{document}

\title{First-principle Investigation of Boron Nitride Nanobelt}

\author{Leonardo S. Barbosa}
\email{leofis94@gmail.com}
\affiliation{Institute of Physics, University of Brasília, Brasília, DF, Brazil}

\author{Bruna C. C. de Almeida}
\affiliation{Institute of Physics, University of Brasília, Brasília, DF, Brazil}

\author{Edvan Moreira}
\affiliation{Physics Department, State University of Maranhão (UEMA), São Luís, MA, Brazil.}

\author{David L. Azevedo}
\affiliation{Institute of Physics, University of Brasília, Brasília, DF, Brazil}

%%%%%%%%%%%%%%%%%%%%%%%%%%%%%%%%%%%%%%%%%%%%%%%%%%%%%%%%%%%%%%%%%%%%%%%%%%%%%%%%%%%%%
\begin{abstract}
In this paper, we report a new boron nitride molecular structure called BN-nanobelt, an inorganic analog of (12)cyclophenacene synthesized in 2017. An extensive investigation using Density Functional Theory (DFT) and Quantum Molecular Dynamics (QMD) calculations showed that BN-nanobelt is a structurally and thermally stable molecule with all positive vibrational frequencies. BN-nanobelt behaves as an insulator, and it absorbs in the ultraviolet region, suggesting a potential application as a UV detector. All results presented in this paper indicate structural stability and the possibility of its synthesis. We hope that the proposed BN-nanobelt structure could stimulate further experimental investigations on its synthesis and bring potential novel technological applications.
\end{abstract}
%%%%%%%%%%%%%%%%%%%%%%%%%%%%%%%%%%%%%%%%%%%%%%%%%%%%%%%%%%%%%%%%%%%%%%%%%%%%%%%%%%%%%

\maketitle

\date{\today}

\section{Introduction}
Over the last half-century, theoretical and experimental research of a new kind of organic molecules has been made. Heilbronner \cite{heilbronner} was the first who introduce a new class of aromatic molecules called carbon nanobelt. His studies dealt with the orbital structure in the hypothetical (n)cyclacenes~\cite{eisenberg}. These aromatic belts predict radially oriented p orbitals and interesting photoluminescence properties confirmed by experiment \cite{Povie172,2019nanobelt}. Since Heilbronner inaugurates aromatic carbon nanobelt investigations, other groups proposed new aromatic nanobelts of (n)cyclacene type~\cite{CORY,Kohnke,Itami}, however, they have faced some difficulties in its synthesis due to strain and high reactivity \cite{LU2017}. Carbon nanobelt follows the same symmetry nomenclature patterns of the nanotubes (armchair and zigzag) and consists of a loop of fully fused benzene rings. Nanobelt of (n)cyclacene type follows a zigzag pattern and (n)cyclophenacene type follows armchair pattern. V\"{o}gtle proposed and attempted the synthesis of a new kind of armchair carbon nanobelt called V\"{o}gtle's belts \cite{vogtle,LU2017}, unfortunately, the synthesis was unsuccessful. Other groups \cite{herges,iyoda2012,bodwell,scott} pursued other ways to make armchair nanobelts, but they have not succeeded. 
 
\begin{figure}[H]
    \centering
    \includegraphics[scale=0.4]{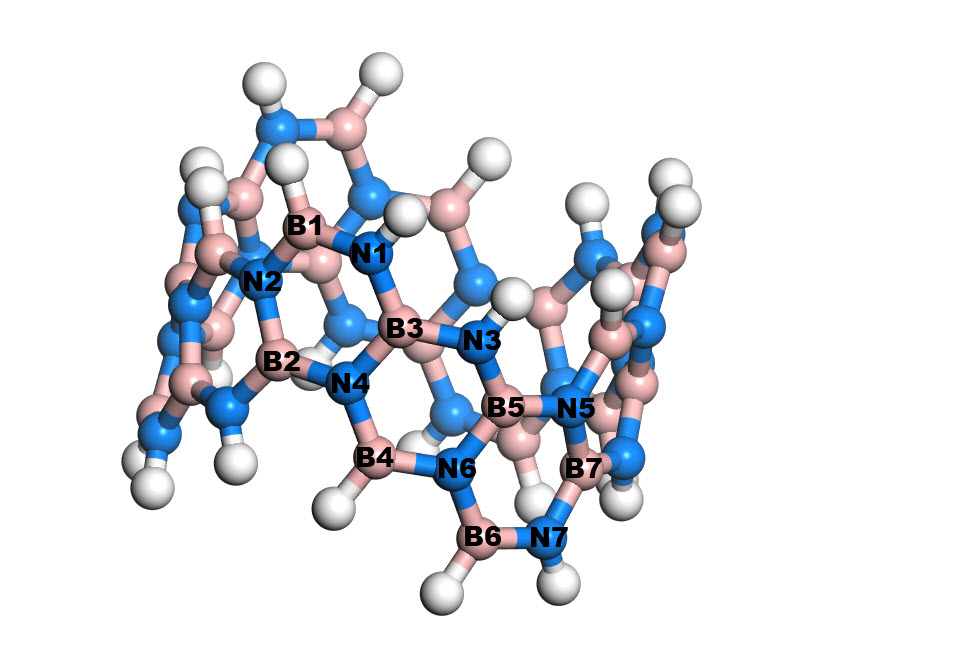}
    \caption{Geometric structure of BN-nanobelt. }
    \label{fig1}
\end{figure}

In 2017, a new armchair carbon nanobelt called (12)cyclophenacene was synthesized by Itami, and Segawa's group \cite{Povie172} which represent segments of an armchair nanotube (6,6). This achievement opens a new way of materials synthesis. Inspired by Itami and Segawa's achievement, Kwan Yin Cheung \textit{et al} \cite{2019nanobelt} recently reported the synthesis of two carbon nanobelts, which represent sidewall segments of an armchair and chiral carbon nanotubes (CNTs), (12,12) and (18,12), respectively. In 2021 an achievement in materials science has been reported, the synthesis of a zigzag carbon nanobelt \cite{zigzag-synthesis}. Zigzag carbon nanobelts are considered the most difficult to synthesize between these two symmetry types (Armchair and Zigzag). This difficulty is due to the high reactivity of zigzag nanobelts \cite{LU2017,zigzag-synthesis,leo-artigo}. Most recently, nitrogen-doped nanobelts were synthesized \cite{dopednanobelt}. Based on these recent synthetic achievements, we can see the importance of the conjugated carbon nanobelts. In this work, encouraged by these discoveries, we propose a boron nitride nanobelt $(B_{24}N_{24}H_{24})$ with the same geometry as (12)cyclophenacene as shown in Figure~\ref{fig1}.
 
To describe this new structure, an extensive study was carried out using the Density Functional Theory method \cite{Kohn1964,KohnSham1965} by calculating electronic, optical, thermodynamic properties and quantum dynamics. It is hoped that BN-nanobelt investigation could serve as seeds to open a new way to synthesize boron nitride nanotubes. This study is organized as follows: In Section 2, the computational methodologies used to obtain all the properties of the proposed structure. Section 3 discusses the results obtained, including structural and thermal stability, electronic properties, optical absorption, vibrational properties, thermodynamics potentials, and quantum dynamics simulation. In Section 4 are presented the concluding remarks.

%\paragraph{Installation} If the document class \emph{elsarticle} is not available on your computer, you can download and install the system package \emph{texlive-publishers} (Linux) or install the \LaTeX\ package \emph{elsarticle} using the package manager of your \TeX\ installation, which is typically \TeX\ Live or Mik\TeX.

%\paragraph{Usage} Once the package is properly installed, you can use the document class \emph{elsarticle} to create a manuscript. Please make sure that your manuscript follows the guidelines in the Guide for Authors of the relevant journal. It is not necessary to typeset your manuscript in exactly the same way as an article, unless you are submitting to a camera-ready copy (CRC) journal.

%\paragraph{Functionality} The Elsevier article class is based on the standard article class and supports almost all of the functionality of that class. In addition, it features commands and options to format the
%\begin{itemize}
%\item document style
%\item baselineskip
%\item front matter
%\item keywords and MSC codes
%\item theorems, definitions and proofs
%\item lables of enumerations
%\item citation style and labeling.
%\end{itemize}

\section{Methodologies}

BN-nanobelt was modeling through molecular software DMol$^{3}$ \cite{delleybasis,dmol3}. The full geometry optimization was performed using LDA-PWC \cite{Pwc} and GGA-PBE \cite{PBE1996} functional. We used the Double Numerical Plus Polarization (DNP) basis set, which includes a polarization p function on all hydrogen atoms. Furthermore, for an appropriate convergence, we considered a maximum energy change below 1.0x$10^{-5} Ha$, maximum force of 2.0x$10^{-3}\frac{Ha}{\text{\AA}}$ and a maximum displacement not exceeding 5.0x$10^{-3}\text{\AA}$. In all calculations, we employed the DFT semi-core pseudopotentials (DSPP) \cite{dspp} method to core electrons treatment which rewrites the core electrons by a single effective potential. 

After obtaining the optimized molecular nanostructure through the DMol$^{3}$ code, the infrared and Raman frequencies were evaluated together with the Raman spectrum considering only the GGA-PBE functional due to the geometry optimization criteria were more stringent than other functionals. The vibrational properties to the
boron nitride nanobelt ($B_{24}N_{24}H_{24}$) will be presented and compared with other literature results.
 
The optical absorption was evaluated for both LDA-PWC and GGA-PBE functionals using the DNP basis set and considering the $80$ first excited states. In addition, the COSMO solvation model was applied to evaluate solvent effects on the optical absorption  \cite{cosmo}. The dichloromethane solvent, with dielectric constant $\varepsilon$=9.08, was considered. The adiabatic local exchange functional approximation (ALDA) \cite{Delleytddft} was used with time-dependent functional theory (TDDFT) \cite{TDDFT}.   
The BN-nanobelt thermodynamic properties were calculated through only PBE functional and DNP basis sets. All calculations are done at $0$\thinspace K than to estimate thermodynamic properties at finite temperature, and it was necessary considering translational, rotational, and vibrational components to determine enthalpy, entropy, free energy, and heat capacity at constant pressure \cite{mopac,wilsonarticle}. 

To verify the stability of the BN-nanobelt structure under high temperatures, we carried out fully atomistic quantum molecular dynamics (QMD) simulations using SIESTA (Spanish Initiative for Electronic Simulations with Thousands of Atoms)~\cite{Siesta} in an NVT ensemble with Nose thermostat, with temperatures ranging from 1000 to 5000\thinspace K, in steps of 500\thinspace K, time step 1\thinspace fs, and the production time of 2.0\thinspace ps. To fine-tuning the temperature of rupture, we carried out a QMD from 3000 to 4000\thinspace K range in steps of 100\thinspace K. For Boron, Nitrogen, and Hydrogen atoms, we replaced the atomic core for a non-relativistic pseudopotential of Ceperly-Alder~\cite{Ceperley1980} without core-correction, where the electronic ground state configuration [2s$^{2}$2p$^{1}$], [2s$^{2}$2p$^{3}$] and [1s$^{1}$] were used respectively. A mesh-cutoff of 50 Rydberg ($\sim$680\thinspace eV) for 3D grid calculation for some integrals and the representation of charge densities and potentials were used.

\section{Results and discussion}

\subsection{Structural stability and electronic properties}
The total energy and the binding energy (energy needed to dissociate the molecular structure) for BN-nanobelt were found and compared with (12)cyclophenacene~\cite{Povie172,leo-artigo} molecule using the GGA-PBE approach as shown in Table \ref{tab1}. This table shows that BN-nanobelt's total energy is smaller than (12)cyclophenacene, about $1.45$ times. This difference can be explained when we look for the electronic Hamiltonian, which describes each one of the molecular systems. BN-nanobelt has more electrons than (12)cyclophenacene, which decreases the system energy. Binding energy results show that BN binding energy is lesser when compared to (12)cyclophenacene; this indicates that carbon-carbon bonds are stronger and requires higher energies to dissociate than the boron-nitrogen bonds. Analyzing the results, we can see the good electronic stability of BN-nanobelt.

In order to support our calculations, we have calculated the borazine (B$_3$H$_6$N$_3$) bond length and bond angle comparing with the available experimental results. The mean borazine bond length is 1.429(1)\thinspace \r{A}~\cite{borazine} while our results are 1.423 and 1.437\thinspace \r{A} considering PWC and PBE functional respectively in the DNP basis set. The borazine mean interbond angles of the six‐membered ring are 117.1(1) degrees at the boron atoms and 122.9(1) degrees at the nitrogen atoms~\cite{borazine} while our results show interbond angles are 116.8 degrees at the boron atoms and 123.1 degrees at the nitrogen atoms (mean values) for both PWC and PBE functional. These results are consistent with the BN-nanobelt bonds length and angle (see Table~\ref{tab2}). It seems that BN-nanobelt consists of fully borazine rings. 

The calculated bond lengths for the BN-nanobelt were also carried out through PWC and PBE and are compared in Table \ref{tab2}. Our results for the calculated BN-nanobelt bond lengths are in good  agreement with the two-dimensional hexagonal boron nitride (h-BN) bond length $1.45\thinspace \text{\AA}$~\cite{h-bn} and boron nitride nanotubes (BNNTs) bond lengths which is between $1.437\thinspace \text{\AA}$ and $1.454\thinspace \text{\AA}$~\cite{chen2015nanotubes}.

\begin{table}[htbp]
\centering
\caption{Total Energy and Binding Energy (in Hartree) of (12)cyclophenacene and BN-nanobelts considering PWC and PBE functionals and DNP basis sets.}
\label{tab1}
\resizebox{\textwidth}{!}{%
\begin{tabular}{lcclcc}
\hline
\textbf{Molecule} & \multicolumn{2}{c}{\textbf{(12)cyclophenacene}} &  & \multicolumn{2}{c}{\textbf{BN-nanobelt}} \\ \cline{2-3} \cline{5-6} 
\textbf{Functional} & \textbf{PWC} & \textbf{PBE} &  & \textbf{PWC} & \textbf{PBE} \\ \hline
\textbf{Basis set} & \textbf{DNP} & \textbf{DNP} &  & \textbf{DNP} & \textbf{DNP} \\
\textbf{Total Energy} & \multicolumn{1}{r}{-1827.47} & \multicolumn{1}{r}{-1841.83} &  & \multicolumn{1}{r}{-1910.95} & \multicolumn{1}{r}{-1925.47} \\
\textbf{Binding Energy} & \multicolumn{1}{r}{-17.56} & \multicolumn{1}{r}{-15.79} &  & \multicolumn{1}{r}{-15.81} & \multicolumn{1}{r}{-14.11} \\ \hline
\end{tabular}%
}
\end{table}

\begin{table}[htbp]
\centering
\caption{Calculated bond length (\AA) and bond angle (degree) of BN-nanobelt considering PWC and PBE functional in DNP basis set.}
\label{tab2}
\resizebox{\textwidth}{!}{%
\begin{tabular}{lrrlrr}
\hline
\multicolumn{6}{c}{\textbf{BN-nanobelt}} \\ \hline
\multicolumn{1}{c}{\textbf{Bond (\AA)}} & \multicolumn{1}{c}{\textbf{PWC}} & \multicolumn{1}{c}{\textbf{PBE}} & \multicolumn{1}{c}{\textbf{Bond Angle (degree)}} & \multicolumn{1}{c}{\textbf{PWC}} & \multicolumn{1}{c}{\textbf{PBE}} \\ \hline
\textbf{B1-N1} & 1.413 & 1.426 & \textbf{B1-N1-B3} & 121.011 & 121.004 \\
\textbf{B1-N2} & 1.444 & 1.459 & \textbf{B3-N4-B2} & 117.736 & 117.555 \\
\textbf{B2-N2} & 1.453 & 1.468 & \textbf{B2-N2-B1} & 119.863 & 119.771 \\
\textbf{B2-N4} & 1.446 & 1.461 & \textbf{N2-B1-N1} & 118.048 & 117.949 \\
\textbf{B3-N1} & 1.437 & 1.451 & \textbf{N1-B3-N4} & 117.985 & 117.963 \\
\textbf{B3-N3} & 1.425 & 1.439 & \textbf{N4-B2-N2} & 119.446 & 119.459 \\
\textbf{B3-N4} & 1.451 & 1.465 & \textbf{B3-N3-B5} & 119.493 & 119.744 \\
\textbf{B4-N4} & 1.436 & 1.453 & \textbf{B5-N6-B4} & 119.562 & 118.696 \\
\textbf{B4-N6} & 1.427 & 1.443 & \textbf{B4-N4-B3} & 118.492 & 118.135 \\
\textbf{B5-N3} & 1.428 & 1.443 & \textbf{N4-B3-N3} & 118.830 & 118.878 \\
\textbf{B5-N5} & 1.446 & 1.461 & \textbf{N3-B5-N6} & 117.061 & 116.613 \\
\textbf{B5-N6} & 1.453 & 1.468 & \textbf{N6-B4-N4} & 120.160 & 120.164 \\
\textbf{B6-N6} & 1.444 & 1.459 & \textbf{B5-N5-B7} & 117.692 & 117.556 \\
\textbf{B6-N7} & 1.413 & 1.426 & \textbf{B7-N7-B6} & 120.933 & 120.965 \\
\textbf{B7-N5} & 1.451 & 1.465 & \textbf{B6-N6-B5} & 119.846 & 119.718 \\
\textbf{B7-N7} & 1.437 & 1.451 & \textbf{N6-B5-N5} & 119.444 & 119.556 \\ \hline
\end{tabular}%
}
\end{table}

%\begin{table}[H]
%\centering
%\caption{Calculated bond length of BN-nanobelt considering PWC and PBE functional in DNP basis set.}
%\label{tab2}
%\begin{tabular}{|l|r|r|}
%\hline
%\multicolumn{1}{|c|}{\textbf{Bonds (in A)}} & \multicolumn{1}{c|}{\textbf{\begin{tabular}[c]{@{}c@{}}BN-nanobelt\\ PWC/DNP\end{tabular}}} & \multicolumn{1}{c|}{\textbf{\begin{tabular}[c]{@{}c@{}}BN-nanobelt\\ PBE/DNP\end{tabular}}} \\ \hline
%\textbf{B1-N1} & 1.413 & 1.426 \\ \hline
%\textbf{B1-N2} & 1.444 & 1.459 \\ \hline
%\textbf{B2-N2} & 1.453 & 1.468 \\ \hline
%\textbf{B2-N4} & 1.446 & 1.461 \\ \hline
%\textbf{B3-N1} & 1.437 & 1.451 \\ \hline
%\textbf{B3-N3} & 1.425 & 1.439 \\ \hline
%\textbf{B3-N4} & 1.451 & 1.465 \\ \hline
%\textbf{B4-N4} & 1.436 & 1.453 \\ \hline
%\textbf{B4-N6} & 1.427 & 1.443 \\ \hline
%\textbf{B5-N3} & 1.428 & 1.443 \\ \hline
%\textbf{B5-N5} & 1.446 & 1.461 \\ \hline
%\textbf{B5-N6} & 1.453 & 1.468 \\ \hline
%\textbf{B6-N6} & 1.444 & 1.459 \\ \hline
%\textbf{B6-N7} & 1.413 & 1.426 \\ \hline
%\textbf{B7-N5} & 1.451 & 1.465 \\ \hline
%\textbf{B7-N7} & 1.437 & 1.451 \\ \hline
%\end{tabular}
%\end{table}

%\subsection{Electronic properties}
The HOMO (highest occupied molecular orbital), LUMO (Lowest unoccupied molecular orbital), and GAP (HOMO-LUMO) energies are shown in Table \ref{tab3} for both molecular systems. The hardness ($\eta$) can be approximated using Koopman's theorem \cite{KOOPMANS1934104} as follows \cite{hardness1,parr}:
  
  \begin{align}
      \eta = \frac{\left (\epsilon LUMO -\epsilon HOMO \right )}{2}
  \end{align}
  
\begin{figure}[htbp]
    \centering
    \includegraphics[scale=0.3]{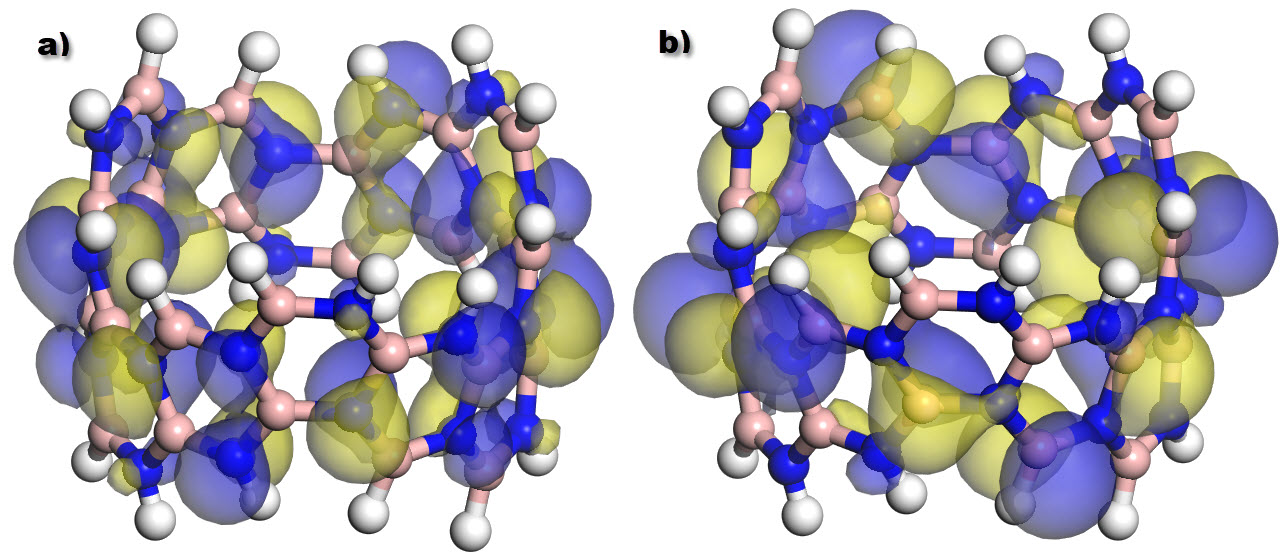}
    \caption{BN-nanobelt frontier molecular orbitals calculated at PBE/DNP  (Isovalue 0.015). a) BN-nanobelt HOMO representation and b) BN-nanobelt LUMO representation. Blue and yellow colors represent the positive (module square of real part of wavefunction) and negative (module square of imaginary part of wavefunction),
respectively. The boron, nitrogen and hydrogen atoms are illustrated by pink, blue, and white spheres, respectively }
    \label{fig3}
\end{figure}

 Hard molecules have a large HOMO-LUMO GAP, and soft molecules have a small HOMO-LUMO gap \cite{hardness2}. The hardness has an important role in chemical reactivity wherein high HOMO-LUMO GAP is responsible for high kinetic stability, and small HOMO-LUMO GAP indicates low chemical stability \cite{hardness}, this suggests that the BN-nanobelt is a very stable structure. From Table \ref{tab3} the GAP results suggest that BN-nanobelt presents an insulator property. This insulator behavior was expected since other boron nitride structures as nanotubes (BNNTs) are insulators with a wide bandgap (5-6\thinspace eV) \cite{chen2015nanotubes,bngap,bngap2,BN-optical2}. The representation of the frontier molecular orbitals is exhibited in Figure~\ref{fig3}. From Figure~\ref{fig3}, it is clear the HOMO preference in nitrogen atoms. This preference is due to electronegativity since the nitrogen atom is more electronegative than the boron atom.
 
To investigate the electronic properties of the proposed BN-nanobelt, we calculated the gap energy, the total density of states (DOS), and the partial density of states (PDOS) after the optimization of geometry using LDA-PWC and GGA-PBE functionals. The DOS of a molecular system describes the number of states allowed to be occupied at each energy range.

\begin{table}[htbp]
\centering
\caption{HOMO, LUMO, GAP and Hardness ($\eta$) energies (in eV) of (12)cyclophenacene~\cite{leo-artigo} and BN-nanobelts considering PBE and PWC functionals with DNP basis sets.}
\label{tab3}
\resizebox{\textwidth}{!}{%
\begin{tabular}{lcclcc}
\hline
\textbf{Molecule} & \multicolumn{2}{l}{\textbf{(12)cyclophenacene}} &  & \multicolumn{2}{l}{\textbf{BN-nanobelt}} \\ \cline{2-3} \cline{5-6} 
\textbf{Functional} & \textbf{PWC} & \textbf{PBE} &  & \textbf{PWC} & \textbf{PBE} \\ \hline
\textbf{Basis set} & \textbf{DNP} & \textbf{DNP} &  & \textbf{DNP} & \textbf{DNP} \\
\textbf{$\epsilon$ HOMO} & -4.84 & -4.64 &  & -6.14 & -5.92 \\
\textbf{$\epsilon$LUMO} & -3.04 & -2.85 &  & -1.58 & -1.30 \\
\textbf{GAP} & 1.80 & 1.79 &  & 4.56 & 4.62 \\
\textbf{Hardness ($\eta$)} & 0.90 & 0.89 &  & 2.28 & 2.31 \\ \hline
\end{tabular}%
}
\end{table}
 
\begin{figure}[htbp]
    \centering
    \includegraphics[scale=0.5]{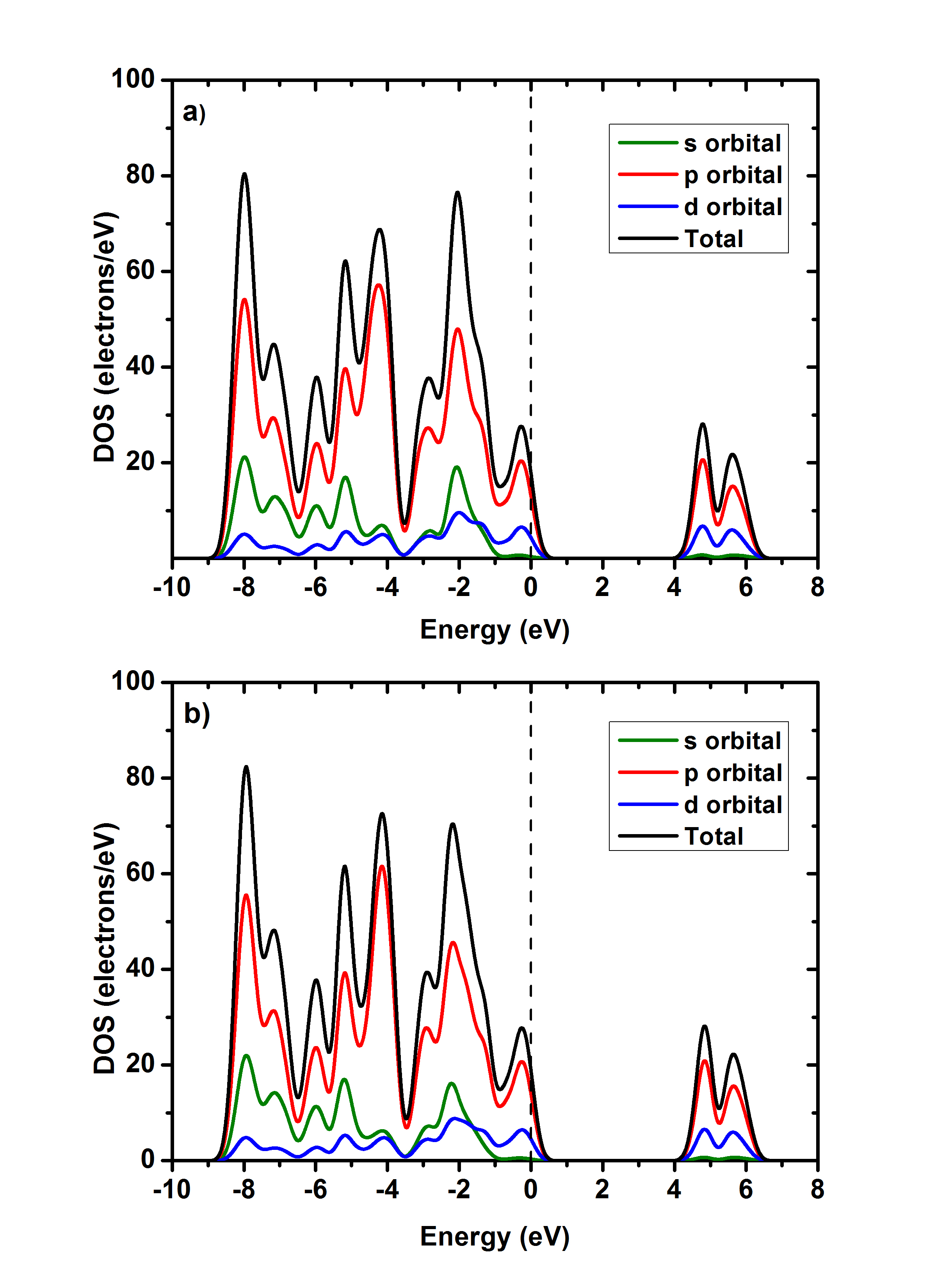}
    \caption{BN-nanobelt total DOS and PDOS a) LDA-PWC, b) GGA-PBE. }
    \label{fig4}
\end{figure}

%\begin{table}[H]
%\centering
%\caption{HOMO, LUMO, GAP and Hardness ($\eta$) energies (in eV) of (12)cyclophenacene and BN-nanobelts considering PBE and PWC functionals with DNP basis sets.} 
%\label{tab3}
%\resizebox{\textwidth}{!}{%
%\begin{tabular}{|l|r|r|r|r|}
%\hline
%\textbf{Molecule} & \multicolumn{2}{l|}{\textbf{(12)cyclophenacene}} & \multicolumn{2}{l|}{\textbf{BN-nanobelt}} \\ \hline
%\textbf{Functional} & \multicolumn{1}{c|}{\textbf{PWC}} & \multicolumn{1}{c|}{\textbf{PBE}} & \multicolumn{1}{c|}{\textbf{PWC}} & \multicolumn{1}{c|}{\textbf{PBE}} \\ \hline
%\textbf{Basis set} & \multicolumn{1}{c|}{\textbf{DNP}} & \multicolumn{1}{c|}{\textbf{DNP}} & \multicolumn{1}{c|}{\textbf{DNP}} & \multicolumn{1}{c|}{\textbf{DNP}} \\ \hline
%\textbf{$\epsilon$HOMO} & -4.84 & -4.64 & -6,14 & -5,92 \\ \hline
%\textbf{$\epsilon$LUMO} & -3.04 & -2.85 & -1,58 & -1,30 \\ \hline
%\textbf{GAP} & 1.80 & 1.79 & 4.56 & 4.62 \\ \hline
%\textbf{Hardness ($\eta$)} & 0.90 & 0.89 & 2.28 & 2.31 \\ \hline
%\end{tabular}%
%}
%\end{table}
From total DOS, Figure~\ref{fig4} can be concluded the insulation character of BN-nanobelt through the analysis of Fermi level whereas the Fermi level is taken to be $0$\thinspace eV. We find that p orbital has a great contribution to the DOS around the Fermi level.
From PDOS it can be inferred that the atomic p orbitals contribute more in the valence region and in the conduction region the p and d orbitals contribute predominantly. 

\subsection{Vibrational properties}

The DMol$^{3}$ code generate a Hessian matrix, a matrix of Cartesian
second derivatives, using this to perform a frequency calculation.
Harmonic vibrational frequencies are computed by diagonalizing the
mass-weighted second-derivative matrix,
\textbf{F}~\cite{Wilson1980}. The elements of \textbf{F} are given
by Eq.~\ref{fij}:

\begin{equation}
\begin{array}{l}
 F_{ij}  = \frac{1}{{\sqrt {m_i m_j } }}\frac{{\partial ^2 E}}{{\partial q_i \partial q_j
 }},
 \end{array}
\label{fij}
\end{equation}
being that, $q_{i}$ and $q_{j}$ represent two Cartesian coordinates
of atoms $i$ and $j$, and $m_{i}$ and $m_{j}$ are the masses of the
atoms. The square roots of the eigenvalues of \textbf{F} are the
harmonic frequencies.

To carry out the calculations of the infrared (IR) intensities of a
given mode, this can be evaluated as a square of all transition
moments of this mode and expressed in terms of the A (atomic polar
tensors) matrix and eigenvectors of the mass-weighted Hessian,
according to the Eq.~\ref{ii}~\cite{Wilson1980}:

\begin{equation}
\begin{array}{l}
I_i  = \left( {\sum\limits_{j,k} {F'_{i,j} A_{j,k} } } \right)^2.
 \end{array}
\label{ii}
\end{equation}
$F'$ are eigenvectors of the normal mode, i.

The Raman spectrum is based on the Raman effect of inelastic
scattering of monochromatic light, resulting in the energy of
incident photons being shifted up or down, employed to study the
vibrational, rotational, and other low-frequency modes. The energy
shift is defined by the vibrational frequency and the proportion of
the inelastically scattered light is defined by the spatial
derivatives of the macroscopic polarization \cite{Porezag1996}, as
also described in previous works to others types of systems
following this same scheme
\cite{Moreira,Moreira1,jefs2013,Moreira2,Moreira3,wilsonarticle}.
However, after calculating Raman activities, the DMol$^{3}$ code can
display corresponding Raman cross sections (intensity) for the
Stokes component of the \emph{i}$^{th}$ mode for a given experiment
incident light frequency and temperature using the Eq.~\ref{Iram}
\cite{Porezag1996}:

\begin{equation}
\begin{array}{l}
\frac{{d\sigma }}{{d\Omega }} = \frac{{\left( {2\pi v_s } \right)^4
}}{{c^4 }}\frac{{h\left( {n_i  + 1} \right)}}{{8\pi ^2 v_c
}}\frac{{I^{Ram} }}{{45}},
\end{array}
\label{Iram}
\end{equation}
being that, $v_{s}$ is the frequency of scattered light, $c$ is the
speed of light, $h$ is Planck constant, $n_{i}$ is the Bose-Einstein
statistical factor, and $I^{Ram}$ is the Raman activity of the given
mode. The parameter $v_s$ can be obtained from the frequency of
incident light $v_0$, i.e., $v_s=v_0-v_i$, we can calculate Raman
intensities in experimental conditions $T$ and $v_0$.

The symmetry group of the molecule of boron nitride nanobelt
($B_{24}N_{24}H_{24}$) is $S_6$ with about 106 active modes for both
IR and Raman. The calculated frequencies for the nanobelt molecule
through GGA approach are shown in Table~\ref{freq}, with some
frequencies close to theoretical calculations and experimental
results, despite being related to periodic structures of the
single-wall boron nitride
nanotubes~\cite{Zhukovskii,Hamdi,Erba,Wirtz,Hasi,Arenal,Nemanich}.
Looking to the IR theoretical frequencies, according to the
Table~\ref{freq}, in the low-frequency region from 26 to
100\thinspace cm$^{-1}$, characterized by the stretching movement of N-H
and B-H bonds, bending of H-B-N, N-B-N, and H-N-B bonds, and
the torsional motion of B-N-B-N, H-B-N-B, H-N-B-N bonds. In the region
from 100 to 211\thinspace cm$^{-1}$ being related to the stretching
movement of N-H and B-H bonds, bending of N-B-N, H-N-B and H-B-N
bonds, the torsional motion of H-B-N-B, B-N-B-N, and H-N-B-N bonds, and
the out-of-plane torsional motion of N-N-N-B and B-B-B-N bonds. Between
frequencies 223 and 390\thinspace cm$^{-1}$ it also includes a
stretching movement of B-N bonds, in addition to those mentioned in
the previous region. The region between 448 and 1200\thinspace
cm$^{-1}$ (with Raman and IR active modes~\cite{Wirtz}), the
vibrational modes are represented by: stretching movement of N-H,
B-H and B-N bonds, bending of N-B-N, H-N-B, and H-B-N bonds,
the torsional motion of H-B-N-B, B-N-B-N, H-N-B-N, and N-B-N-B
(additional mode) bonds, and out-of-plane torsional motion of
N-N-N-B and B-B-B-N bonds. In the 1224-1500\thinspace cm$^{-1}$
range at medium frequencies (with Raman and IR active
modes~\cite{Wirtz}), the vibrational modes are related to the
stretching movement of N-H, B-H, and B-N bonds, bending of N-B-N,
H-N-B and H-B-N bonds, the torsional motion of H-B-N-B, B-N-B-N,
H-N-B-N, and N-B-N-B bonds, and out-of-plane torsional motion of
N-N-N-B and B-B-B-N bonds, with emphasis on the stronger stretching
movement of B-N bonds. In the region from higher frequencies,
between 1500 and 2610\thinspace cm$^{-1}$, approximately, the modes
are assigned to the very strong stretching movement of H-B bonds,
bending of N-B-N, H-N-B, and H-B-N bonds, the torsional motion of
H-B-N-B, B-N-B-N, H-N-B-N, and N-B-N-B bonds, and out-of-plane
the torsional motion of N-N-N-B and B-B-B-N bonds. From 3500\thinspace
cm$^{-1}$, the vibrational modes are related with a very strong
stretching movement of H-N bonds, bending of N-B-N, H-N-B, and H-B-N
bonds, and torsional motion of H-B-N-B, B-N-B-N, H-N-B-N, and
N-B-N-B bonds.

\begin{figure}[htbp]
\centerline{\includegraphics[width=1.00\textwidth]{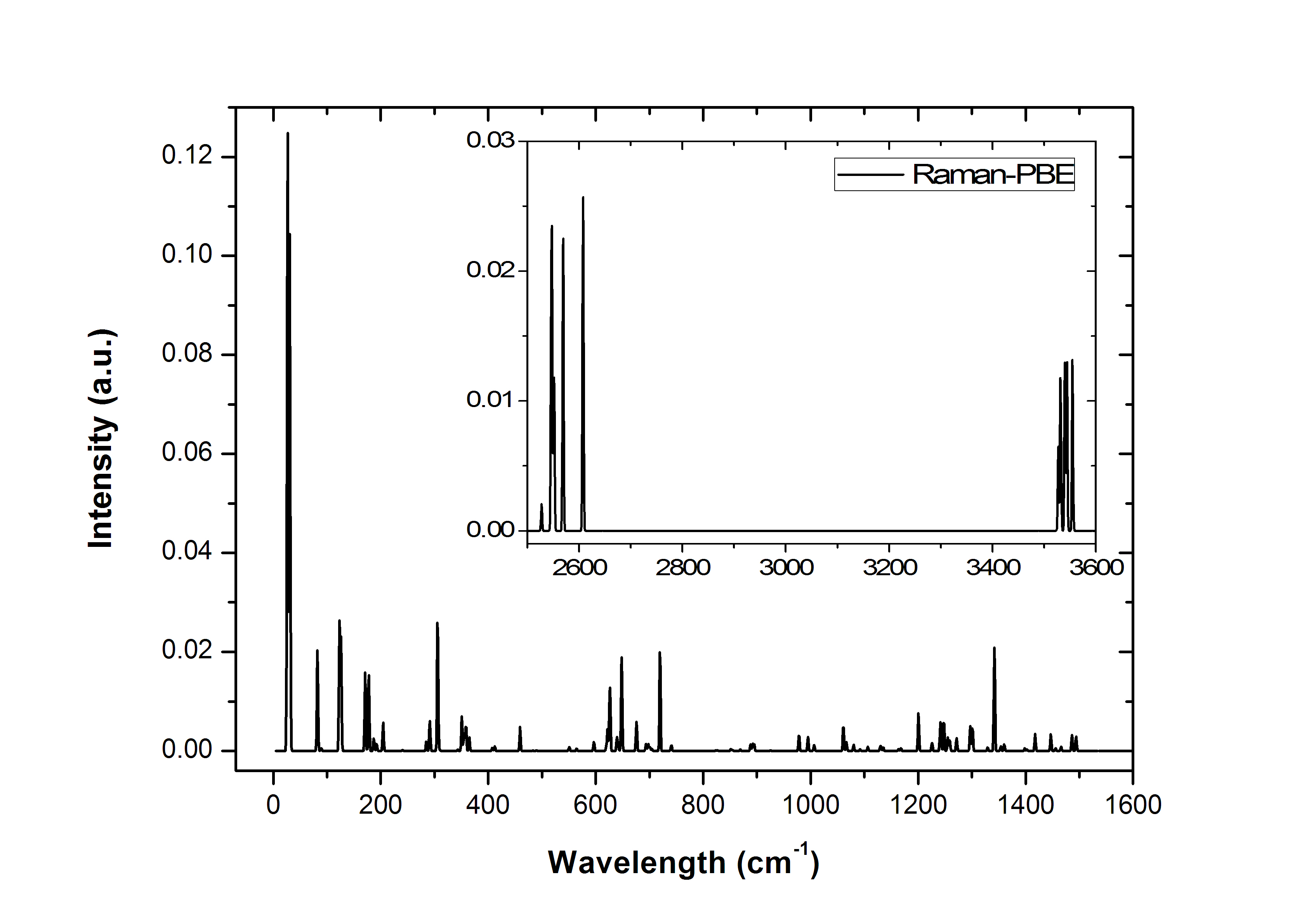}}
\caption{Raman spectrum of nanobelt in the 0-3600\thinspace
cm$^{-1}$ range using the GGA-PBE exchange-correlation
functional.}\label{raman}
\end{figure}

Figure~\ref{raman} presents the calculated Raman scattering
spectrum, using the GGA-PBE functional. In the 0-1600\thinspace
cm$^{-1}$ range, the most intense peak at 26.8\thinspace cm$^{-1}$,
in the low-frequency regime, is assigned to the movements of
stretching of N-H bonds, bending of H-B-N, N-B-N, and H-N-B bonds,
and torsional motion of B-N-B-N bonds. The second most intense peak appears at 123.5\thinspace cm$^{-1}$ is related to the stretching movement of N-H and B-H bonds, bending of N-B-N, H-N-B, and H-B-N bonds, the torsional motion of H-B-N-B, B-N-B-N, and H-N-B-N bonds, and out-of-plane torsional motion of N-N-N-B and B-B-B-N bonds. The third most intense peak occurs at 305.9\thinspace cm$^{-1}$, corresponding to the stretching movement of N-H, B-H, and B-N bonds, bending of N-B-N, H-N-B, and H-B-N bonds, the torsional motion of H-B-N-B, B-N-B-N, and H-N-B-N bonds, and out-of-plane torsional motion of N-N-N-B and B-B-B-N bonds. There is a fourth intense peak in 1360.3\thinspace cm$^{-1}$ (1364.7~\cite{Zhukovskii}, 1376~\cite{Hamdi}, 1365~\cite{Arenal}, and 1366\thinspace cm$^{-1}$~\cite{Nemanich}, associated with tangential modes at high frequencies for periodic structures of h-BN, which is the main peak in spectra) that can be mainly associated with a stronger stretching movement of B-N bonds and a bending of N-B-N, H-N-B, and H-B-N bonds.
Smaller peaks can be seen up to 1600\thinspace cm$^{-1}$ related
with the vibrational modes in the same range as the IR active modes.
In the 2500-3600\thinspace cm$^{-1}$ range, for higher frequencies,
the most intense peak occurs at 2607.5\thinspace cm$^{-1}$ mainly
associated with a very strong stretching movement of H-B bonds.

%\begin{landscape}
\begin{table}[htbp]
\begin{center}
\caption{IR and Raman assignment table of BN-nanobelt}
%\vskip1truecm
\tiny
\begin{tabular}{llllllllllllllll}
\hline

\multicolumn{1}{c|}{N} & \multicolumn{1}{c}{$k(cm^{-1})$} & \multicolumn{1}{c}{IR} & \multicolumn{1}{c|}{Raman} & \multicolumn{1}{c|}{N} & \multicolumn{1}{c}{$k(cm^{-1})$} & \multicolumn{1}{c}{IR} & \multicolumn{1}{c|}{Raman} & \multicolumn{1}{c|}{N} & \multicolumn{1}{c}{$k(cm^{-1})$} & \multicolumn{1}{c}{IR} & \multicolumn{1}{c|}{Raman} & \multicolumn{1}{c|}{N} & \multicolumn{1}{c}{$k(cm^{-1})$} & \multicolumn{1}{c}{IR} & \multicolumn{1}{c}{Raman}\\
\hline

\multicolumn{1}{c|}{1} & \multicolumn{1}{c}{26.8} & \multicolumn{1}{c}{N} & \multicolumn{1}{c|}{Y} & \multicolumn{1}{c|}{54} & \multicolumn{1}{c}{512.3} & \multicolumn{1}{c}{Y} & \multicolumn{1}{c|}{N} & \multicolumn{1}{c|}{107} & \multicolumn{1}{c}{887.1} & \multicolumn{1}{c}{Y} & \multicolumn{1}{c|}{N} & \multicolumn{1}{c|}{160} & \multicolumn{1}{c}{1297.0} & \multicolumn{1}{c}{N} & \multicolumn{1}{c}{Y}\\
\multicolumn{1}{c|}{2} & \multicolumn{1}{c}{30.9} & \multicolumn{1}{c}{N} & \multicolumn{1}{c|}{Y} & \multicolumn{1}{c|}{55} & \multicolumn{1}{c}{550.5} & \multicolumn{1}{c}{N} & \multicolumn{1}{c|}{Y} & \multicolumn{1}{c|}{108} & \multicolumn{1}{c}{887.2} & \multicolumn{1}{c}{N} & \multicolumn{1}{c|}{Y} & \multicolumn{1}{c|}{161} & \multicolumn{1}{c}{1301.1} & \multicolumn{1}{c}{N} & \multicolumn{1}{c}{N}\\
\multicolumn{1}{c|}{3} & \multicolumn{1}{c}{35.3} & \multicolumn{1}{c}{Y} & \multicolumn{1}{c|}{N} & \multicolumn{1}{c|}{56} & \multicolumn{1}{c}{551.3} & \multicolumn{1}{c}{N} & \multicolumn{1}{c|}{Y} & \multicolumn{1}{c|}{109} & \multicolumn{1}{c}{889.4} & \multicolumn{1}{c}{Y} & \multicolumn{1}{c|}{N} & \multicolumn{1}{c|}{162} & \multicolumn{1}{c}{1301.8} & \multicolumn{1}{c}{Y} & \multicolumn{1}{c}{Y}\\
\multicolumn{1}{c|}{4} & \multicolumn{1}{c}{47.0} & \multicolumn{1}{c}{Y} & \multicolumn{1}{c|}{N} & \multicolumn{1}{c|}{57} & \multicolumn{1}{c}{553.6} & \multicolumn{1}{c}{Y} & \multicolumn{1}{c|}{N} & \multicolumn{1}{c|}{110} & \multicolumn{1}{c}{891.8} & \multicolumn{1}{c}{N} & \multicolumn{1}{c|}{Y} & \multicolumn{1}{c|}{163} & \multicolumn{1}{c}{1329.2} & \multicolumn{1}{c}{N} & \multicolumn{1}{c}{Y}\\
\multicolumn{1}{c|}{5} & \multicolumn{1}{c}{75.9} & \multicolumn{1}{c}{Y} & \multicolumn{1}{c|}{N} & \multicolumn{1}{c|}{58} & \multicolumn{1}{c}{559.0} & \multicolumn{1}{c}{Y} & \multicolumn{1}{c|}{N} & \multicolumn{1}{c|}{111} & \multicolumn{1}{c}{895.0} & \multicolumn{1}{c}{N} & \multicolumn{1}{c|}{Y} & \multicolumn{1}{c|}{164} & \multicolumn{1}{c}{1342.0} & \multicolumn{1}{c}{N} & \multicolumn{1}{c}{Y}\\
\multicolumn{1}{c|}{6} & \multicolumn{1}{c}{78.2} & \multicolumn{1}{c}{Y} & \multicolumn{1}{c|}{N} & \multicolumn{1}{c|}{59} & \multicolumn{1}{c}{561.8} & \multicolumn{1}{c}{Y} & \multicolumn{1}{c|}{N} & \multicolumn{1}{c|}{112} & \multicolumn{1}{c}{901.8} & \multicolumn{1}{c}{Y} & \multicolumn{1}{c|}{N} & \multicolumn{1}{c|}{165} & \multicolumn{1}{c}{1347.1} & \multicolumn{1}{c}{Y} & \multicolumn{1}{c}{N}\\
\multicolumn{1}{c|}{7} & \multicolumn{1}{c}{82.2} & \multicolumn{1}{c}{N} & \multicolumn{1}{c|}{Y} & \multicolumn{1}{c|}{60} & \multicolumn{1}{c}{564.6} & \multicolumn{1}{c}{N} & \multicolumn{1}{c|}{Y} & \multicolumn{1}{c|}{113} & \multicolumn{1}{c}{904.3} & \multicolumn{1}{c}{Y} & \multicolumn{1}{c|}{N} & \multicolumn{1}{c|}{166} & \multicolumn{1}{c}{1349.9} & \multicolumn{1}{c}{Y} & \multicolumn{1}{c}{N}\\
\multicolumn{1}{c|}{8} & \multicolumn{1}{c}{89.3} & \multicolumn{1}{c}{N} & \multicolumn{1}{c|}{Y} & \multicolumn{1}{c|}{61} & \multicolumn{1}{c}{596.6} & \multicolumn{1}{c}{N} & \multicolumn{1}{c|}{Y} & \multicolumn{1}{c|}{114} & \multicolumn{1}{c}{925.7} & \multicolumn{1}{c}{N} & \multicolumn{1}{c|}{N} & \multicolumn{1}{c|}{167} & \multicolumn{1}{c}{1352.5} & \multicolumn{1}{c}{Y} & \multicolumn{1}{c}{N}\\
\multicolumn{1}{c|}{9} & \multicolumn{1}{c}{123.5} & \multicolumn{1}{c}{N} & \multicolumn{1}{c|}{Y} & \multicolumn{1}{c|}{62} & \multicolumn{1}{c}{601.0} & \multicolumn{1}{c}{Y} & \multicolumn{1}{c|}{N} & \multicolumn{1}{c|}{115} & \multicolumn{1}{c}{978.1} & \multicolumn{1}{c}{N} & \multicolumn{1}{c|}{Y} & \multicolumn{1}{c|}{168} & \multicolumn{1}{c}{1353.8} & \multicolumn{1}{c}{N} & \multicolumn{1}{c}{Y}\\
\multicolumn{1}{c|}{10} & \multicolumn{1}{c}{126.1} & \multicolumn{1}{c}{N} & \multicolumn{1}{c|}{N} & \multicolumn{1}{c|}{63} & \multicolumn{1}{c}{604.1} & \multicolumn{1}{c}{Y} & \multicolumn{1}{c|}{N} & \multicolumn{1}{c|}{116} & \multicolumn{1}{c}{981.8} & \multicolumn{1}{c}{Y} & \multicolumn{1}{c|}{N} & \multicolumn{1}{c|}{169} & \multicolumn{1}{c}{1360.3} & \multicolumn{1}{c}{N} & \multicolumn{1}{c}{Y}\\
\multicolumn{1}{c|}{11} & \multicolumn{1}{c}{134.0} & \multicolumn{1}{c}{Y} & \multicolumn{1}{c|}{N} & \multicolumn{1}{c|}{64} & \multicolumn{1}{c}{619.3} & \multicolumn{1}{c}{N} & \multicolumn{1}{c|}{Y} & \multicolumn{1}{c|}{117} & \multicolumn{1}{c}{995.2} & \multicolumn{1}{c}{N} & \multicolumn{1}{c|}{Y} & \multicolumn{1}{c|}{170} & \multicolumn{1}{c}{1391.4} & \multicolumn{1}{c}{Y} & \multicolumn{1}{c}{N}\\
\multicolumn{1}{c|}{12} & \multicolumn{1}{c}{135.2} & \multicolumn{1}{c}{Y} & \multicolumn{1}{c|}{N} & \multicolumn{1}{c|}{65} & \multicolumn{1}{c}{622.0} & \multicolumn{1}{c}{N} & \multicolumn{1}{c|}{Y} & \multicolumn{1}{c|}{118} & \multicolumn{1}{c}{997.0} & \multicolumn{1}{c}{Y} & \multicolumn{1}{c|}{Y} & \multicolumn{1}{c|}{171} & \multicolumn{1}{c}{1395.0} & \multicolumn{1}{c}{Y} & \multicolumn{1}{c}{N}\\
\multicolumn{1}{c|}{13} & \multicolumn{1}{c}{157.7} & \multicolumn{1}{c}{Y} & \multicolumn{1}{c|}{N} & \multicolumn{1}{c|}{66} & \multicolumn{1}{c}{626.2} & \multicolumn{1}{c}{N} & \multicolumn{1}{c|}{Y} & \multicolumn{1}{c|}{119} & \multicolumn{1}{c}{1005.0} & \multicolumn{1}{c}{Y} & \multicolumn{1}{c|}{Y} & \multicolumn{1}{c|}{172} & \multicolumn{1}{c}{1398.2} & \multicolumn{1}{c}{N} & \multicolumn{1}{c}{Y}\\
\multicolumn{1}{c|}{14} & \multicolumn{1}{c}{171.0} & \multicolumn{1}{c}{N} & \multicolumn{1}{c|}{N} & \multicolumn{1}{c|}{67} & \multicolumn{1}{c}{632.7} & \multicolumn{1}{c}{Y} & \multicolumn{1}{c|}{N} & \multicolumn{1}{c|}{120} & \multicolumn{1}{c}{1006.4} & \multicolumn{1}{c}{N} & \multicolumn{1}{c|}{Y} & \multicolumn{1}{c|}{173} & \multicolumn{1}{c}{1402.6} & \multicolumn{1}{c}{N} & \multicolumn{1}{c}{Y}\\
\multicolumn{1}{c|}{15} & \multicolumn{1}{c}{177.6} & \multicolumn{1}{c}{N} & \multicolumn{1}{c|}{N} & \multicolumn{1}{c|}{68} & \multicolumn{1}{c}{633.5} & \multicolumn{1}{c}{Y} & \multicolumn{1}{c|}{N} & \multicolumn{1}{c|}{121} & \multicolumn{1}{c}{1060.7} & \multicolumn{1}{c}{N} & \multicolumn{1}{c|}{Y} & \multicolumn{1}{c|}{174} & \multicolumn{1}{c}{1416.0} & \multicolumn{1}{c}{Y} & \multicolumn{1}{c}{N}\\
\multicolumn{1}{c|}{16} & \multicolumn{1}{c}{186.6} & \multicolumn{1}{c}{Y} & \multicolumn{1}{c|}{N} & \multicolumn{1}{c|}{69} & \multicolumn{1}{c}{639.7} & \multicolumn{1}{c}{N} & \multicolumn{1}{c|}{Y} & \multicolumn{1}{c|}{122} & \multicolumn{1}{c}{1061.8} & \multicolumn{1}{c}{Y} & \multicolumn{1}{c|}{N} & \multicolumn{1}{c|}{175} & \multicolumn{1}{c}{1417.5} & \multicolumn{1}{c}{N} & \multicolumn{1}{c}{Y}\\
\multicolumn{1}{c|}{17} & \multicolumn{1}{c}{187.0} & \multicolumn{1}{c}{N} & \multicolumn{1}{c|}{N} & \multicolumn{1}{c|}{70} & \multicolumn{1}{c}{640.5} & \multicolumn{1}{c}{Y} & \multicolumn{1}{c|}{N} & \multicolumn{1}{c|}{123} & \multicolumn{1}{c}{1065.5} & \multicolumn{1}{c}{Y} & \multicolumn{1}{c|}{N} & \multicolumn{1}{c|}{176} & \multicolumn{1}{c}{1429.2} & \multicolumn{1}{c}{Y} & \multicolumn{1}{c}{N}\\
\multicolumn{1}{c|}{18} & \multicolumn{1}{c}{192.1} & \multicolumn{1}{c}{N} & \multicolumn{1}{c|}{N} & \multicolumn{1}{c|}{71} & \multicolumn{1}{c}{642.2} & \multicolumn{1}{c}{N} & \multicolumn{1}{c|}{N} & \multicolumn{1}{c|}{124} & \multicolumn{1}{c}{1066.3} & \multicolumn{1}{c}{N} & \multicolumn{1}{c|}{Y} & \multicolumn{1}{c|}{177} & \multicolumn{1}{c}{1437.3} & \multicolumn{1}{c}{Y} & \multicolumn{1}{c}{N}\\
\multicolumn{1}{c|}{19} & \multicolumn{1}{c}{192.2} & \multicolumn{1}{c}{Y} & \multicolumn{1}{c|}{N} & \multicolumn{1}{c|}{72} & \multicolumn{1}{c}{642.7} & \multicolumn{1}{c}{Y} & \multicolumn{1}{c|}{Y} & \multicolumn{1}{c|}{125} & \multicolumn{1}{c}{1077.3} & \multicolumn{1}{c}{Y} & \multicolumn{1}{c|}{N} & \multicolumn{1}{c|}{178} & \multicolumn{1}{c}{1447.0} & \multicolumn{1}{c}{N} & \multicolumn{1}{c}{Y}\\
\multicolumn{1}{c|}{20} & \multicolumn{1}{c}{204.7} & \multicolumn{1}{c}{N} & \multicolumn{1}{c|}{Y} & \multicolumn{1}{c|}{73} & \multicolumn{1}{c}{648.3} & \multicolumn{1}{c}{N} & \multicolumn{1}{c|}{Y} & \multicolumn{1}{c|}{126} & \multicolumn{1}{c}{1079.6} & \multicolumn{1}{c}{N} & \multicolumn{1}{c|}{Y} & \multicolumn{1}{c|}{179} & \multicolumn{1}{c}{1455.7} & \multicolumn{1}{c}{N} & \multicolumn{1}{c}{Y}\\
\multicolumn{1}{c|}{21} & \multicolumn{1}{c}{211.3} & \multicolumn{1}{c}{Y} & \multicolumn{1}{c|}{N} & \multicolumn{1}{c|}{74} & \multicolumn{1}{c}{656.7} & \multicolumn{1}{c}{Y} & \multicolumn{1}{c|}{N} & \multicolumn{1}{c|}{127} & \multicolumn{1}{c}{1080.1} & \multicolumn{1}{c}{N} & \multicolumn{1}{c|}{Y} & \multicolumn{1}{c|}{180} & \multicolumn{1}{c}{1463.4} & \multicolumn{1}{c}{Y} & \multicolumn{1}{c}{N}\\
\multicolumn{1}{c|}{22} & \multicolumn{1}{c}{223.8} & \multicolumn{1}{c}{Y} & \multicolumn{1}{c|}{N} & \multicolumn{1}{c|}{75} & \multicolumn{1}{c}{658.7} & \multicolumn{1}{c}{Y} & \multicolumn{1}{c|}{N} & \multicolumn{1}{c|}{128} & \multicolumn{1}{c}{1082.4} & \multicolumn{1}{c}{Y} & \multicolumn{1}{c|}{N} & \multicolumn{1}{c|}{181} & \multicolumn{1}{c}{1466.0} & \multicolumn{1}{c}{N} & \multicolumn{1}{c}{N}\\
\multicolumn{1}{c|}{23} & \multicolumn{1}{c}{227.2} & \multicolumn{1}{c}{Y} & \multicolumn{1}{c|}{N} & \multicolumn{1}{c|}{76} & \multicolumn{1}{c}{673.0} & \multicolumn{1}{c}{Y} & \multicolumn{1}{c|}{N} & \multicolumn{1}{c|}{129} & \multicolumn{1}{c}{1091.9} & \multicolumn{1}{c}{N} & \multicolumn{1}{c|}{Y} & \multicolumn{1}{c|}{182} & \multicolumn{1}{c}{1467.0} & \multicolumn{1}{c}{Y} & \multicolumn{1}{c}{N}\\
\multicolumn{1}{c|}{24} & \multicolumn{1}{c}{240.5} & \multicolumn{1}{c}{N} & \multicolumn{1}{c|}{Y} & \multicolumn{1}{c|}{77} & \multicolumn{1}{c}{675.7} & \multicolumn{1}{c}{N} & \multicolumn{1}{c|}{Y} & \multicolumn{1}{c|}{130} & \multicolumn{1}{c}{1095.2} & \multicolumn{1}{c}{Y} & \multicolumn{1}{c|}{N} & \multicolumn{1}{c|}{183} & \multicolumn{1}{c}{1478.8} & \multicolumn{1}{c}{Y} & \multicolumn{1}{c}{N}\\
\multicolumn{1}{c|}{25} & \multicolumn{1}{c}{273.8} & \multicolumn{1}{c}{Y} & \multicolumn{1}{c|}{N} & \multicolumn{1}{c|}{78} & \multicolumn{1}{c}{675.9} & \multicolumn{1}{c}{N} & \multicolumn{1}{c|}{Y} & \multicolumn{1}{c|}{131} & \multicolumn{1}{c}{1103.0} & \multicolumn{1}{c}{Y} & \multicolumn{1}{c|}{N} & \multicolumn{1}{c|}{184} & \multicolumn{1}{c}{1486.1} & \multicolumn{1}{c}{N} & \multicolumn{1}{c}{Y}\\
\multicolumn{1}{c|}{26} & \multicolumn{1}{c}{285.0} & \multicolumn{1}{c}{N} & \multicolumn{1}{c|}{Y} & \multicolumn{1}{c|}{79} & \multicolumn{1}{c}{689.5} & \multicolumn{1}{c}{Y} & \multicolumn{1}{c|}{N} & \multicolumn{1}{c|}{132} & \multicolumn{1}{c}{1106.3} & \multicolumn{1}{c}{N} & \multicolumn{1}{c|}{Y} & \multicolumn{1}{c|}{185} & \multicolumn{1}{c}{1493.8} & \multicolumn{1}{c}{N} & \multicolumn{1}{c}{Y}\\
\multicolumn{1}{c|}{27} & \multicolumn{1}{c}{291.3} & \multicolumn{1}{c}{Y} & \multicolumn{1}{c|}{N} & \multicolumn{1}{c|}{80} & \multicolumn{1}{c}{692.7} & \multicolumn{1}{c}{Y} & \multicolumn{1}{c|}{N} & \multicolumn{1}{c|}{133} & \multicolumn{1}{c}{1124.3} & \multicolumn{1}{c}{Y} & \multicolumn{1}{c|}{N} & \multicolumn{1}{c|}{186} & \multicolumn{1}{c}{1497.8} & \multicolumn{1}{c}{Y} & \multicolumn{1}{c}{N}\\
\multicolumn{1}{c|}{28} & \multicolumn{1}{c}{291.4} & \multicolumn{1}{c}{N} & \multicolumn{1}{c|}{Y} & \multicolumn{1}{c|}{81} & \multicolumn{1}{c}{693.0} & \multicolumn{1}{c}{N} & \multicolumn{1}{c|}{Y} & \multicolumn{1}{c|}{134} & \multicolumn{1}{c}{1129.2} & \multicolumn{1}{c}{N} & \multicolumn{1}{c|}{Y} & \multicolumn{1}{c|}{187} & \multicolumn{1}{c}{2527.44} & \multicolumn{1}{c}{N} & \multicolumn{1}{c}{Y}\\
\multicolumn{1}{c|}{29} & \multicolumn{1}{c}{294.7} & \multicolumn{1}{c}{Y} & \multicolumn{1}{c|}{N} & \multicolumn{1}{c|}{82} & \multicolumn{1}{c}{693.7} & \multicolumn{1}{c}{N} & \multicolumn{1}{c|}{Y} & \multicolumn{1}{c|}{135} & \multicolumn{1}{c}{1130.7} & \multicolumn{1}{c}{N} & \multicolumn{1}{c|}{Y} & \multicolumn{1}{c|}{188} & \multicolumn{1}{c}{2527.48} & \multicolumn{1}{c}{Y} & \multicolumn{1}{c}{Y}\\
\multicolumn{1}{c|}{30} & \multicolumn{1}{c}{305.9} & \multicolumn{1}{c}{N} & \multicolumn{1}{c|}{Y} & \multicolumn{1}{c|}{83} & \multicolumn{1}{c}{696.3} & \multicolumn{1}{c}{Y} & \multicolumn{1}{c|}{N} & \multicolumn{1}{c|}{136} & \multicolumn{1}{c}{1135.0} & \multicolumn{1}{c}{N} & \multicolumn{1}{c|}{Y} & \multicolumn{1}{c|}{189} & \multicolumn{1}{c}{2545.9} & \multicolumn{1}{c}{N} & \multicolumn{1}{c}{Y}\\
\multicolumn{1}{c|}{31} & \multicolumn{1}{c}{310.1} & \multicolumn{1}{c}{N} & \multicolumn{1}{c|}{N} & \multicolumn{1}{c|}{84} & \multicolumn{1}{c}{697.7} & \multicolumn{1}{c}{N} & \multicolumn{1}{c|}{Y} & \multicolumn{1}{c|}{137} & \multicolumn{1}{c}{1138.3} & \multicolumn{1}{c}{Y} & \multicolumn{1}{c|}{N} & \multicolumn{1}{c|}{190} & \multicolumn{1}{c}{2546.4} & \multicolumn{1}{c}{Y} & \multicolumn{1}{c}{Y}\\
\multicolumn{1}{c|}{32} & \multicolumn{1}{c}{313.8} & \multicolumn{1}{c}{Y} & \multicolumn{1}{c|}{N} & \multicolumn{1}{c|}{85} & \multicolumn{1}{c}{700.7} & \multicolumn{1}{c}{N} & \multicolumn{1}{c|}{Y} & \multicolumn{1}{c|}{138} & \multicolumn{1}{c}{1139.8} & \multicolumn{1}{c}{Y} & \multicolumn{1}{c|}{N} & \multicolumn{1}{c|}{191} & \multicolumn{1}{c}{2547.3} & \multicolumn{1}{c}{Y} & \multicolumn{1}{c}{Y}\\
\multicolumn{1}{c|}{33} & \multicolumn{1}{c}{320.7} & \multicolumn{1}{c}{Y} & \multicolumn{1}{c|}{N} & \multicolumn{1}{c|}{86} & \multicolumn{1}{c}{704.3} & \multicolumn{1}{c}{N} & \multicolumn{1}{c|}{Y} & \multicolumn{1}{c|}{139} & \multicolumn{1}{c}{1157.1} & \multicolumn{1}{c}{Y} & \multicolumn{1}{c|}{N} & \multicolumn{1}{c|}{192} & \multicolumn{1}{c}{2547.6} & \multicolumn{1}{c}{N} & \multicolumn{1}{c}{N}\\
\multicolumn{1}{c|}{34} & \multicolumn{1}{c}{327.5} & \multicolumn{1}{c}{Y} & \multicolumn{1}{c|}{N} & \multicolumn{1}{c|}{87} & \multicolumn{1}{c}{707.0} & \multicolumn{1}{c}{Y} & \multicolumn{1}{c|}{N} & \multicolumn{1}{c|}{140} & \multicolumn{1}{c}{1163.5} & \multicolumn{1}{c}{N} & \multicolumn{1}{c|}{Y} & \multicolumn{1}{c|}{193} & \multicolumn{1}{c}{2551.6} & \multicolumn{1}{c}{N} & \multicolumn{1}{c}{Y}\\
\multicolumn{1}{c|}{35} & \multicolumn{1}{c}{343.3} & \multicolumn{1}{c}{N} & \multicolumn{1}{c|}{N} & \multicolumn{1}{c|}{88} & \multicolumn{1}{c}{710.5} & \multicolumn{1}{c}{Y} & \multicolumn{1}{c|}{N} & \multicolumn{1}{c|}{141} & \multicolumn{1}{c}{1168.2} & \multicolumn{1}{c}{N} & \multicolumn{1}{c|}{Y} & \multicolumn{1}{c|}{194} & \multicolumn{1}{c}{2551.7} & \multicolumn{1}{c}{Y} & \multicolumn{1}{c}{N}\\
\multicolumn{1}{c|}{36} & \multicolumn{1}{c}{350.8} & \multicolumn{1}{c}{N} & \multicolumn{1}{c|}{Y} & \multicolumn{1}{c|}{89} & \multicolumn{1}{c}{715.6} & \multicolumn{1}{c}{Y} & \multicolumn{1}{c|}{N} & \multicolumn{1}{c|}{142} & \multicolumn{1}{c}{1183.2} & \multicolumn{1}{c}{Y} & \multicolumn{1}{c|}{N} & \multicolumn{1}{c|}{195} & \multicolumn{1}{c}{2568.9} & \multicolumn{1}{c}{N} & \multicolumn{1}{c}{Y}\\
\multicolumn{1}{c|}{37} & \multicolumn{1}{c}{351.5} & \multicolumn{1}{c}{Y} & \multicolumn{1}{c|}{N} & \multicolumn{1}{c|}{90} & \multicolumn{1}{c}{719.5} & \multicolumn{1}{c}{N} & \multicolumn{1}{c|}{Y} & \multicolumn{1}{c|}{143} & \multicolumn{1}{c}{1184.1} & \multicolumn{1}{c}{Y} & \multicolumn{1}{c|}{N} & \multicolumn{1}{c|}{196} & \multicolumn{1}{c}{2569.2} & \multicolumn{1}{c}{Y} & \multicolumn{1}{c}{Y}\\
\multicolumn{1}{c|}{38} & \multicolumn{1}{c}{354.5} & \multicolumn{1}{c}{Y} & \multicolumn{1}{c|}{N} & \multicolumn{1}{c|}{91} & \multicolumn{1}{c}{724.8} & \multicolumn{1}{c}{Y} & \multicolumn{1}{c|}{N} & \multicolumn{1}{c|}{144} & \multicolumn{1}{c}{1200.5} & \multicolumn{1}{c}{N} & \multicolumn{1}{c|}{Y} & \multicolumn{1}{c|}{197} & \multicolumn{1}{c}{2607.58} & \multicolumn{1}{c}{Y} & \multicolumn{1}{c}{Y}\\
\multicolumn{1}{c|}{39} & \multicolumn{1}{c}{354.7} & \multicolumn{1}{c}{N} & \multicolumn{1}{c|}{Y} & \multicolumn{1}{c|}{92} & \multicolumn{1}{c}{740.5} & \multicolumn{1}{c}{N} & \multicolumn{1}{c|}{Y} & \multicolumn{1}{c|}{145} & \multicolumn{1}{c}{1224.1} & \multicolumn{1}{c}{Y} & \multicolumn{1}{c|}{Y} & \multicolumn{1}{c|}{198} & \multicolumn{1}{c}{2607.58} & \multicolumn{1}{c}{Y} & \multicolumn{1}{c}{Y}\\
\multicolumn{1}{c|}{40} & \multicolumn{1}{c}{356.1} & \multicolumn{1}{c}{N} & \multicolumn{1}{c|}{N} & \multicolumn{1}{c|}{93} & \multicolumn{1}{c}{741.3} & \multicolumn{1}{c}{N} & \multicolumn{1}{c|}{Y} & \multicolumn{1}{c|}{146} & \multicolumn{1}{c}{1225.8} & \multicolumn{1}{c}{N} & \multicolumn{1}{c|}{Y} & \multicolumn{1}{c|}{199} & \multicolumn{1}{c}{3527.03} & \multicolumn{1}{c}{N} & \multicolumn{1}{c}{Y}\\
\multicolumn{1}{c|}{41} & \multicolumn{1}{c}{358.4} & \multicolumn{1}{c}{N} & \multicolumn{1}{c|}{Y} & \multicolumn{1}{c|}{94} & \multicolumn{1}{c}{752.0} & \multicolumn{1}{c}{Y} & \multicolumn{1}{c|}{N} & \multicolumn{1}{c|}{147} & \multicolumn{1}{c}{1238.1} & \multicolumn{1}{c}{Y} & \multicolumn{1}{c|}{Y} & \multicolumn{1}{c|}{200} & \multicolumn{1}{c}{3527.08} & \multicolumn{1}{c}{Y} & \multicolumn{1}{c}{Y}\\
\multicolumn{1}{c|}{42} & \multicolumn{1}{c}{364.6} & \multicolumn{1}{c}{N} & \multicolumn{1}{c|}{N} & \multicolumn{1}{c|}{95} & \multicolumn{1}{c}{755.7} & \multicolumn{1}{c}{Y} & \multicolumn{1}{c|}{N} & \multicolumn{1}{c|}{148} & \multicolumn{1}{c}{1240.7} & \multicolumn{1}{c}{N} & \multicolumn{1}{c|}{Y} & \multicolumn{1}{c|}{201} & \multicolumn{1}{c}{3531.11} & \multicolumn{1}{c}{N} & \multicolumn{1}{c}{Y}\\
\multicolumn{1}{c|}{43} & \multicolumn{1}{c}{367.7} & \multicolumn{1}{c}{Y} & \multicolumn{1}{c|}{Y} & \multicolumn{1}{c|}{96} & \multicolumn{1}{c}{761.8} & \multicolumn{1}{c}{N} & \multicolumn{1}{c|}{N} & \multicolumn{1}{c|}{149} & \multicolumn{1}{c}{1242.0} & \multicolumn{1}{c}{N} & \multicolumn{1}{c|}{N} & \multicolumn{1}{c|}{202} & \multicolumn{1}{c}{3531.16} & \multicolumn{1}{c}{Y} & \multicolumn{1}{c}{Y}\\
\multicolumn{1}{c|}{44} & \multicolumn{1}{c}{386.4} & \multicolumn{1}{c}{Y} & \multicolumn{1}{c|}{N} & \multicolumn{1}{c|}{97} & \multicolumn{1}{c}{796.5} & \multicolumn{1}{c}{Y} & \multicolumn{1}{c|}{N} & \multicolumn{1}{c|}{150} & \multicolumn{1}{c}{1242.4} & \multicolumn{1}{c}{Y} & \multicolumn{1}{c|}{Y} & \multicolumn{1}{c|}{203} & \multicolumn{1}{c}{3532.0} & \multicolumn{1}{c}{N} & \multicolumn{1}{c}{N}\\
\multicolumn{1}{c|}{45} & \multicolumn{1}{c}{389.9} & \multicolumn{1}{c}{Y} & \multicolumn{1}{c|}{N} & \multicolumn{1}{c|}{98} & \multicolumn{1}{c}{823.5} & \multicolumn{1}{c}{N} & \multicolumn{1}{c|}{Y} & \multicolumn{1}{c|}{151} & \multicolumn{1}{c}{1246.4} & \multicolumn{1}{c}{Y} & \multicolumn{1}{c|}{Y} & \multicolumn{1}{c|}{204} & \multicolumn{1}{c}{3532.1} & \multicolumn{1}{c}{Y} & \multicolumn{1}{c}{Y}\\
\multicolumn{1}{c|}{46} & \multicolumn{1}{c}{407.6} & \multicolumn{1}{c}{N} & \multicolumn{1}{c|}{N} & \multicolumn{1}{c|}{99} & \multicolumn{1}{c}{826.6} & \multicolumn{1}{c}{N} & \multicolumn{1}{c|}{Y} & \multicolumn{1}{c|}{152} & \multicolumn{1}{c}{1247.9} & \multicolumn{1}{c}{N} & \multicolumn{1}{c|}{Y} & \multicolumn{1}{c|}{205} & \multicolumn{1}{c}{3540.08} & \multicolumn{1}{c}{Y} & \multicolumn{1}{c}{Y}\\
\multicolumn{1}{c|}{47} & \multicolumn{1}{c}{412.2} & \multicolumn{1}{c}{N} & \multicolumn{1}{c|}{Y} & \multicolumn{1}{c|}{100} & \multicolumn{1}{c}{848.4} & \multicolumn{1}{c}{Y} & \multicolumn{1}{c|}{N} & \multicolumn{1}{c|}{153} & \multicolumn{1}{c}{1250.7} & \multicolumn{1}{c}{Y} & \multicolumn{1}{c|}{N} & \multicolumn{1}{c|}{206} & \multicolumn{1}{c}{3540.09} & \multicolumn{1}{c}{N} & \multicolumn{1}{c}{Y}\\
\multicolumn{1}{c|}{48} & \multicolumn{1}{c}{448.8} & \multicolumn{1}{c}{Y} & \multicolumn{1}{c|}{N} & \multicolumn{1}{c|}{101} & \multicolumn{1}{c}{851.1} & \multicolumn{1}{c}{N} & \multicolumn{1}{c|}{Y} & \multicolumn{1}{c|}{154} & \multicolumn{1}{c}{1255.3} & \multicolumn{1}{c}{N} & \multicolumn{1}{c|}{Y} & \multicolumn{1}{c|}{207} & \multicolumn{1}{c}{3544.10} & \multicolumn{1}{c}{Y} & \multicolumn{1}{c}{Y}\\
\multicolumn{1}{c|}{49} & \multicolumn{1}{c}{451.6} & \multicolumn{1}{c}{Y} & \multicolumn{1}{c|}{N} & \multicolumn{1}{c|}{102} & \multicolumn{1}{c}{851.2} & \multicolumn{1}{c}{Y} & \multicolumn{1}{c|}{Y} & \multicolumn{1}{c|}{155} & \multicolumn{1}{c}{1258.8} & \multicolumn{1}{c}{N} & \multicolumn{1}{c|}{Y} & \multicolumn{1}{c|}{208} & \multicolumn{1}{c}{3544.12} & \multicolumn{1}{c}{N} & \multicolumn{1}{c}{Y}\\
\multicolumn{1}{c|}{50} & \multicolumn{1}{c}{456.7} & \multicolumn{1}{c}{Y} & \multicolumn{1}{c|}{N} & \multicolumn{1}{c|}{103} & \multicolumn{1}{c}{854.4} & \multicolumn{1}{c}{N} & \multicolumn{1}{c|}{Y} & \multicolumn{1}{c|}{156} & \multicolumn{1}{c}{1271.6} & \multicolumn{1}{c}{N} & \multicolumn{1}{c|}{Y} & \multicolumn{1}{c|}{209} & \multicolumn{1}{c}{3554.73} & \multicolumn{1}{c}{Y} & \multicolumn{1}{c}{Y}\\
\multicolumn{1}{c|}{51} & \multicolumn{1}{c}{459.2} & \multicolumn{1}{c}{N} & \multicolumn{1}{c|}{Y} & \multicolumn{1}{c|}{104} & \multicolumn{1}{c}{865.3} & \multicolumn{1}{c}{Y} & \multicolumn{1}{c|}{N} & \multicolumn{1}{c|}{157} & \multicolumn{1}{c}{1273.9} & \multicolumn{1}{c}{Y} & \multicolumn{1}{c|}{N} & \multicolumn{1}{c|}{210} & \multicolumn{1}{c}{3554.74} & \multicolumn{1}{c}{N} & \multicolumn{1}{c}{Y}\\
\multicolumn{1}{c|}{52} & \multicolumn{1}{c}{483.2} & \multicolumn{1}{c}{N} & \multicolumn{1}{c|}{Y} & \multicolumn{1}{c|}{105} & \multicolumn{1}{c}{869.0} & \multicolumn{1}{c}{N} & \multicolumn{1}{c|}{Y} & \multicolumn{1}{c|}{158} & \multicolumn{1}{c}{1277.1} & \multicolumn{1}{c}{Y} & \multicolumn{1}{c|}{N} & \multicolumn{1}{c|}{-} & \multicolumn{1}{c}{-} & \multicolumn{1}{c}{-} & \multicolumn{1}{c}{-}\\
\multicolumn{1}{c|}{53} & \multicolumn{1}{c}{489.6} & \multicolumn{1}{c}{N} & \multicolumn{1}{c|}{Y} & \multicolumn{1}{c|}{106} & \multicolumn{1}{c}{878.3} & \multicolumn{1}{c}{Y} & \multicolumn{1}{c|}{N} & \multicolumn{1}{c|}{159} & \multicolumn{1}{c}{1284.5} & \multicolumn{1}{c}{Y} & \multicolumn{1}{c|}{N} & \multicolumn{1}{c|}{-} & \multicolumn{1}{c}{-} & \multicolumn{1}{c}{-} & \multicolumn{1}{c}{-}\\
\hline \label{freq}
\end{tabular}
\end{center}
\end{table}
%\end{landscape}

\subsection{Optical properties}

\begin{figure}[htbp]
    \centering
    \includegraphics[scale=0.5]{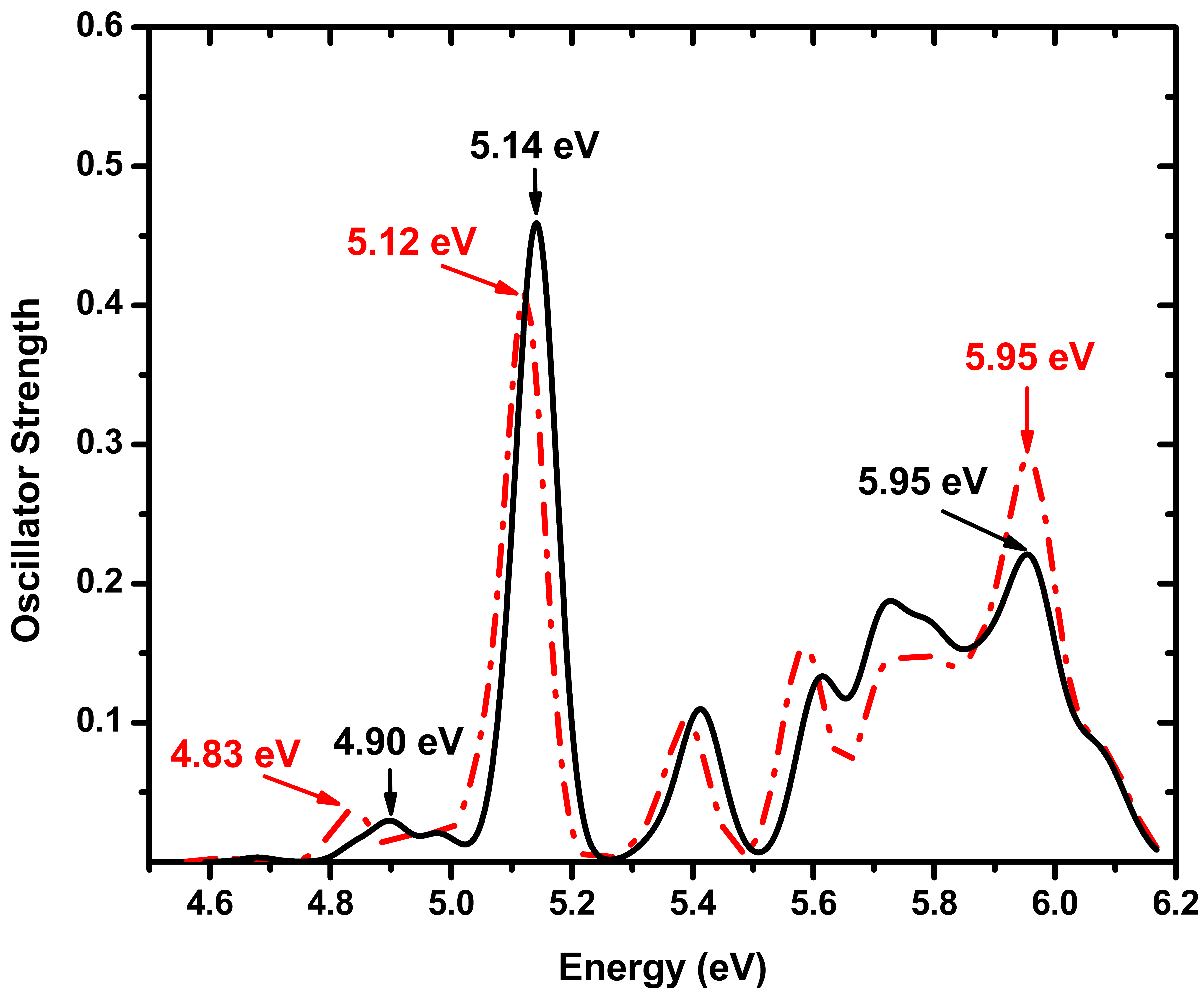}
    \caption{BN-nanobelt absorbance peaks calculations performed through PWC (red dashed line) and PBE (solid black line) functional considering DNP basis set and dichloromethane solvent.}
    \label{fig5}
\end{figure}

Figure \ref{fig5} shows the BN-nanobelt optical absorbance peak. The calculations were performed using LDA-PWC (red dashed line) and GGA-PBE (solid black line) functional considering the DNP basis set and the molecule immersed in dichloromethane. 
The result presents all the excited states. The HOMO-LUMO transitions are the first two small peaks in the energy range between $4.8$\thinspace eV ($258$\thinspace nm) and $5.0$\thinspace eV ($248$\thinspace nm). The maximum absorption peaks are between $5.0$\thinspace eV ($248$\thinspace nm) and $5.3$\thinspace eV ($234$\thinspace nm). The second largest absorption peaks are around at $5.95$\thinspace eV ($208$\thinspace nm). BN-nanobelt absorbance calculated here is in excellent agreement with previous work on boron nitride nanotubes' experimental optic properties, which indicates absorbance peaks at $4.45$, $5.5$, and $6.15$\thinspace eV \cite{BN-optical}.
These optical results suggest that BN-nanobelt has potential for applications in optoelectronic as a UV detector.

\subsection{Thermodynamic properties}

\begin{figure}[htbp]
    \centering
    \includegraphics[scale=0.45]{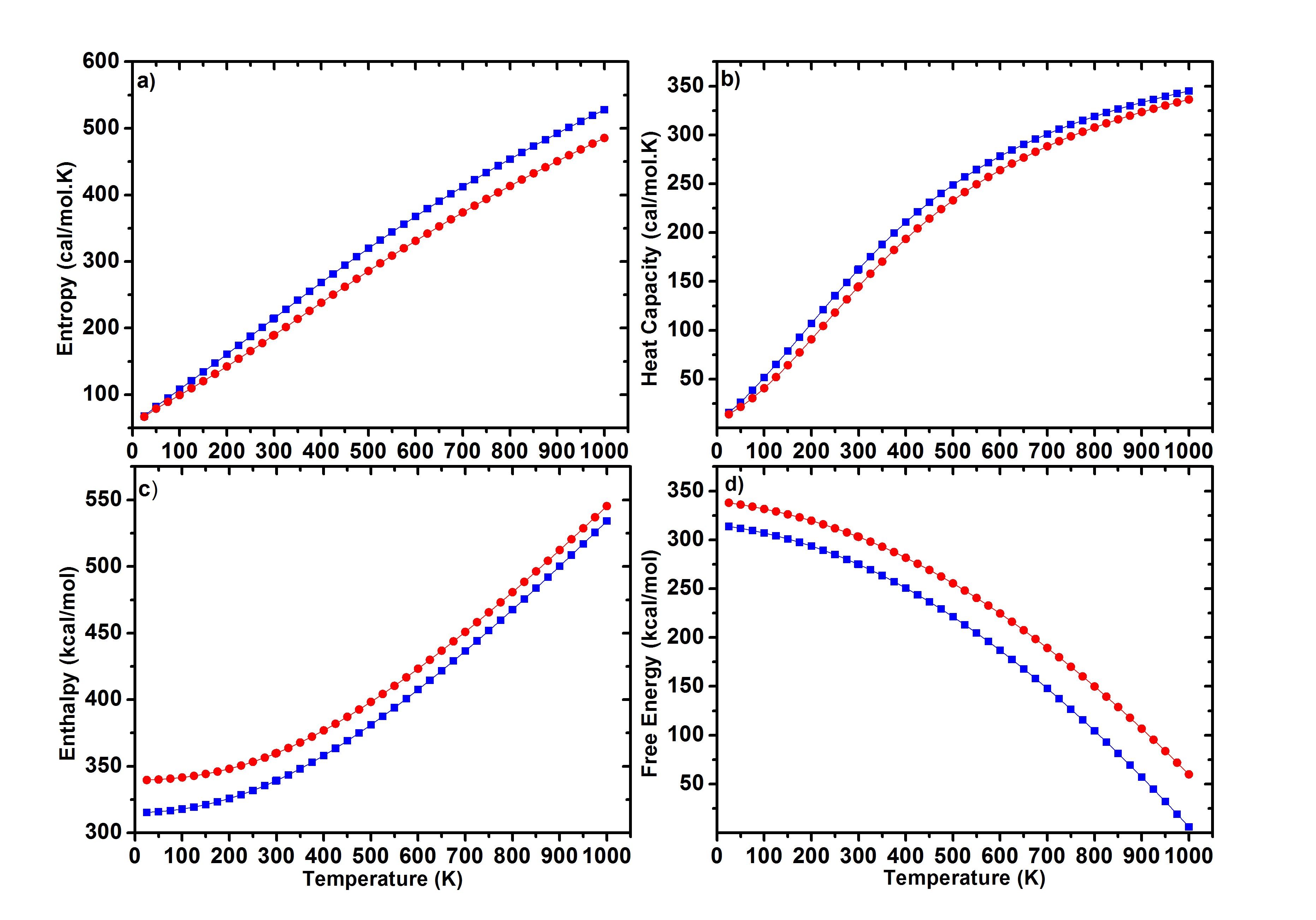}
    \caption{(12)cyclophenacene (circles-red) and BN-nanobelt (square-blue) thermodynamic potentials: a) Entropy, b) Heat Capacity, c) Enthalpy and d) Free energy.}
    \label{fig6}
\end{figure}

In Figure \ref{fig6} it is represented the thermodynamic potentials. The calculated entropy as a function of the temperature in the range of 0-1000\thinspace K, Figure~\ref{fig6} a). The BN-nanobelt has a higher entropy when compared to the (12)cyclophenacene nanobelt. This fact suggests that BN-nanobelt reacts a little bit more to temperature changes than (12)cyclophenacene. 

The Figure~\ref{fig6} b) presents the heat capacity at a constant pressure of both molecular systems. It is possible to view that the curves are well similar; this result implies that BN-nanobelt slightly absorbs more heat than (12)cyclophenacene, this means that BN-nanobelt needs more energy to raise the temperature by 1\thinspace K. BNNTs are known for their high thermal conductivity~\cite{kim2018boron}, so it is expected that BN-nanobelt also presents a high thermal conductivity. 

The enthalpy is represented as a function of the temperature in Figure~\ref{fig6} c). The enthalpy of both molecular systems has an exponential behavior as far as temperature increases. We can see in the temperature range of 0-1000\thinspace K that BN-nanobelt enthalpy (square-blue) is less compared with (12)cyclophenacene; in other words, BN-nanobelt requires less energy to be synthesized than (12)cyclophenacene.   

Finally, Figure~\ref{fig6} d) presents the free energy of both molecules. The results show that in the range of 0-1000\thinspace K, the free energy is positive; in other words, there is not spontaneous formation. For comparison, at $1000$\thinspace K, BN-nanobelt free energy is $6.1$\thinspace kcal/mol, and for (12)cyclophenacene free energy at $1000$\thinspace K is $59.9$\thinspace kcal/mol.  Despite all this, BN-nanobelt presents the lowest free energy compared with (12)cyclophenacene, which indicates a more favorable possibility of its synthesis.

For both molecular structures mentioned here, the entropy and enthalpy increase as temperature increases, indicating that all these reactions are endothermic. Furthermore, BN-nanobelt has a smaller enthalpy and free energy than (12)cyclophenacene, suggesting a potential synthesis. However, we cannot guarantee that synthesis would occur. The process of synthesis is complex due to many factors. Only additional research from an experimental and thermochemical point of view could confirm our findings.

%Although we can not guarantee that synthesis could happen due to
%many factors, just a further and more in-depth research from the experimental
%or thermochemical point of view could confirm our findings

\subsection{Quantum dynamics properties}

Using SIESTA code~\cite{Siesta}, we performed quantum molecular dynamics to analyze the stability of BN-nanobelt in high temperatures. Figure~\ref{fig7} below presents frames for molecular dynamics simulation indicating the frames where this molecular structure starts to break. We observed that BN-nanobelt suffers a rupture at 3000\thinspace K ($\sim$2727\thinspace $^{\circ}$C ), indicating high thermal stability for this structure.

\begin{figure}[htbp]
    \centering
    \includegraphics[scale=0.3]{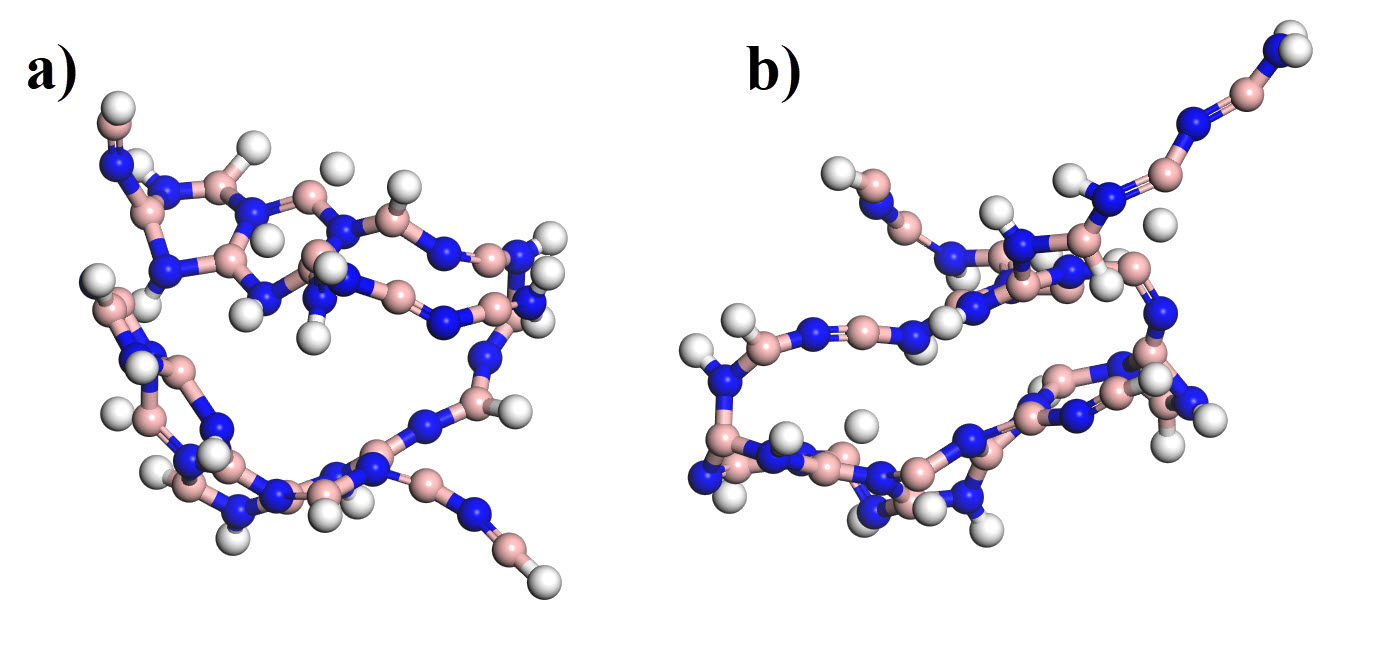}
    \caption{Rupture frames for BN-nanobelt a) frame 1070, b) frame 1474. }
    \label{fig7}
\end{figure}

\section{Conclusions}

In this paper, we proposed the BN-nanobelt. We performed \textit{ab initio} calculations within the DFT approach to calculate electronic, optical, and thermodynamics properties. The BN-nanobelt stability was verified through a vibrational frequencies analysis wherein all frequencies are positives. We verified that BN-nanobelt possesses an insulation character with an estimated gap of around $4.6$\thinspace eV. The hardness analysis for BN-nanobelt showed high chemical stability. The optical absorption spectra revealed that BN-nanobelt is an excellent UV absorber, suggesting that this molecule could be employed as a UV detector. The calculated thermodynamics potentials indicate that BN-nanobelt could also be synthesized. %This work aims to motivate future experiments on boron nitride nanobelt and nanotubes.  
The calculated IR and Raman frequencies show the following assignments of its active modes, stretching, bending, and torsion of the bonds. The calculated Raman spectrum exhibits the most intense peak at 26.8\thinspace cm$^{-1}$, in the low-frequency regime, corresponding to movements of stretching of N-H bonds, bending of H-B-N, N-B-N and H-N-B bonds, and torsional motion of B-N-B-N bonds. Finally, quantum molecular dynamics results indicated that BN-nanobelt presents high stability at a very high temperature (around 3000\thinspace K).

\section*{Acknowledgements}
This study was partially financed by the Brazilian Research Agencies CAPES and CNPq. The authors also thank the Distrito Federal Research Foundation FAPDF Edital 04/2017 for financial and equipment support. E. Moreira acknowledges the support by the Maranhão Research Foundation -- FAPEMA (Universal-01108/19). David L. Azevedo acknowledges the support by the Mato Grosso Research Foundation FAPEMAT for financial support through the Grant PRONEX CNPq/ FAPEMAT 850109/2009.

%\section{Front matter}

%The author names and affiliations could be formatted in two ways:
%\begin{enumerate}[(1)]
%\item Group the authors per affiliation.
%\item Use footnotes to indicate the affiliations.
%\end{enumerate}
%See the front matter of this document for examples. You are recommended to conform your choice to the journal you are submitting to.

%\section{Bibliography styles}

%There are various bibliography styles available. You can select the style of your choice in the preamble of this document. These styles are Elsevier styles based on standard styles like Harvard and Vancouver. Please use Bib\TeX\ to generate your bibliography and include DOIs whenever available.

%Here are two sample references: \cite{Feynman1963118,Dirac1953888}.

%\section*{References}


\begin{thebibliography}{10}
\expandafter\ifx\csname url\endcsname\relax
  \def\url#1{\texttt{#1}}\fi
\expandafter\ifx\csname urlprefix\endcsname\relax\def\urlprefix{URL }\fi
\expandafter\ifx\csname href\endcsname\relax
  \def\href#1#2{#2} \def\path#1{#1}\fi

\bibitem{heilbronner}
E.~Heilbronner, Molecular orbitals in homologen reihen mehrkerniger
  aromatischer kohlenwasserstoffe: I. die eigenwerte yon lcao-mo's in homologen
  reihen, Helvetica Chimica Acta 37~(3) (1954) 921--935.
\newblock \href {http://dx.doi.org/10.1002/hlca.19540370336}
  {\path{doi:10.1002/hlca.19540370336}}.

\bibitem{eisenberg}
D.~Eisenberg, R.~Shenhar, M.~Rabinovitz, Synthetic approaches to aromatic
  belts: building up strain in macrocyclic polyarenes, Chem. Soc. Rev. 39
  (2010) 2879--2890.
\newblock \href {http://dx.doi.org/10.1039/B904087K}
  {\path{doi:10.1039/B904087K}}.

\bibitem{Povie172}
G.~Povie, Y.~Segawa, T.~Nishihara, Y.~Miyauchi, K.~Itami, Synthesis of a carbon
  nanobelt, Science 356~(6334) (2017) 172--175.
\newblock \href {http://dx.doi.org/10.1126/science.aam8158}
  {\path{doi:10.1126/science.aam8158}}.

\bibitem{2019nanobelt}
K.~Y. Cheung, S.~Gui, C.~Deng, H.~Liang, Z.~Xia, Z.~Liu, L.~Chi, Q.~Miao,
  Synthesis of armchair and chiral carbon nanobelts, Chem 5~(4) (2019) 838 --
  847.
\newblock \href
  {http://dx.doi.org/https://doi.org/10.1016/j.chempr.2019.01.004}
  {\path{doi:https://doi.org/10.1016/j.chempr.2019.01.004}}.

\bibitem{CORY}
R.~M. Cory, C.~L. McPhail, A.~J. Dikmans, J.~J. Vittal, Macrocyclic cyclophane
  belts via double diels-alder cycloadditions: Macroannulation of bisdienes by
  bisdienophiles. synthesis of a key precursor to an [8]cyclacene, Tetrahedron
  Letters 37~(12) (1996) 1983 -- 1986.
\newblock \href
  {http://dx.doi.org/https://doi.org/10.1016/0040-4039(96)00263-8}
  {\path{doi:https://doi.org/10.1016/0040-4039(96)00263-8}}.

\bibitem{Kohnke}
F.~H. Kohnke, A.~M.~Z. Slawin, J.~F. Stoddart, D.~J. Williams, Molecular belts
  and collars in the making: A hexaepoxyoctacosahydro[12]cyclacene derivative,
  Angewandte Chemie International Edition in English 26~(9) (1987) 892--894.
\newblock \href {http://dx.doi.org/10.1002/anie.198708921}
  {\path{doi:10.1002/anie.198708921}}.

\bibitem{Itami}
Y.~Segawa, A.~Yagi, H.~Ito, K.~Itami, A theoretical study on the strain energy
  of carbon nanobelts, Organic Letters 18~(6) (2016) 1430--1433.
\newblock \href {http://dx.doi.org/10.1021/acs.orglett.6b00365}
  {\path{doi:10.1021/acs.orglett.6b00365}}.

\bibitem{LU2017}
X.~Lu, J.~Wu, After 60 years of efforts: The chemical synthesis of a carbon
  nanobelt, Chem 2~(5) (2017) 619 -- 620.
\newblock \href
  {http://dx.doi.org/https://doi.org/10.1016/j.chempr.2017.04.012}
  {\path{doi:https://doi.org/10.1016/j.chempr.2017.04.012}}.

\bibitem{vogtle}
F.~Vögtle, A.~Schröder, D.~Karbach, Strategy for the synthesis of tube-shaped
  molecules, Angewandte Chemie International Edition in English 30~(5) (1991)
  575--577.
\newblock \href {http://dx.doi.org/10.1002/anie.199105751}
  {\path{doi:10.1002/anie.199105751}}.

\bibitem{herges}
S.~Kammermeier, P.~G. Jones, R.~Herges, Ring-expanding metathesis of
  tetradehydro-anthracene—synthesis and structure of a tubelike, fully
  conjugated hydrocarbon, Angewandte Chemie International Edition in English
  35~(22) (1996) 2669--2671.
\newblock \href {http://dx.doi.org/10.1002/anie.199626691}
  {\path{doi:10.1002/anie.199626691}}.

\bibitem{iyoda2012}
M.~Iyoda, Y.~Kuwatani, T.~Nishinaga, M.~Takase, T.~Nishiuchi, Conjugated
  Molecular Belts Based on 3D Benzannulene Systems, John Wiley \& Sons, Ltd,
  2011, Ch.~12, pp. 311--342.
\newblock \href {http://dx.doi.org/10.1002/9781118011263.ch12}
  {\path{doi:10.1002/9781118011263.ch12}}.

\bibitem{bodwell}
B.~Merner, L.~Dawe, G.~Bodwell, 1,1,8,8-tetramethyl[8](2,11)teropyrenophane:
  Half of an aromatic belt and a segment of an (8,8) single-walled carbon
  nanotube, Angewandte Chemie International Edition 48~(30) (2009) 5487--5491.
\newblock \href {http://dx.doi.org/10.1002/anie.200806363}
  {\path{doi:10.1002/anie.200806363}}.

\bibitem{scott}
L.~T. Scott, Conjugated belts and nanorings with radially oriented p orbitals,
  Angewandte Chemie International Edition 42~(35) (2003) 4133--4135.
\newblock \href {http://dx.doi.org/10.1002/anie.200301671}
  {\path{doi:10.1002/anie.200301671}}.

\bibitem{zigzag-synthesis}
K.~Y. Cheung, K.~Watanabe, Y.~Segawa, K.~Itami, Synthesis of a zigzag carbon
  nanobelt, Nature Chemistry 13~(3) (2021) 255--259.
\newblock \href {http://dx.doi.org/10.1038/s41557-020-00627-5}
  {\path{doi:10.1038/s41557-020-00627-5}}.

\bibitem{leo-artigo}
L.~S. Barbosa, L.~A. Leal, R.~Gargano, D.~L. Azevedo, Silicon carbide nanobelt:
  A novel molecule with potential technological application, Computational and
  Theoretical Chemistry 1171 (2020) 112645.
\newblock \href
  {http://dx.doi.org/https://doi.org/10.1016/j.comptc.2019.112645}
  {\path{doi:https://doi.org/10.1016/j.comptc.2019.112645}}.

\bibitem{dopednanobelt}
J.~Zhu, Y.~Han, Y.~Ni, G.~Li, J.~Wu, Facile synthesis of nitrogen-doped
  [(6.)m8]ncyclacene carbon nanobelts by a one-pot self-condensation reaction,
  Journal of the American Chemical Society 143~(7) (2021) 2716--2721.
\newblock \href {http://dx.doi.org/10.1021/jacs.1c00409}
  {\path{doi:10.1021/jacs.1c00409}}.

\bibitem{Kohn1964}
P.~Hohenberg, W.~Kohn, Inhomogeneous electron gas, Phys. Rev. 136 (1964)
  B864--B871.
\newblock \href {http://dx.doi.org/10.1103/PhysRev.136.B864}
  {\path{doi:10.1103/PhysRev.136.B864}}.

\bibitem{KohnSham1965}
W.~Kohn, L.~J. Sham, Self-consistent equations including exchange and
  correlation effects, Phys. Rev. 140 (1965) A1133--A1138.
\newblock \href {http://dx.doi.org/10.1103/PhysRev.140.A1133}
  {\path{doi:10.1103/PhysRev.140.A1133}}.

\bibitem{delleybasis}
B.~Delley, An all‐electron numerical method for solving the local density
  functional for polyatomic molecules, The Journal of Chemical Physics 92~(1)
  (1990) 508--517.
\newblock \href {http://dx.doi.org/10.1063/1.458452}
  {\path{doi:10.1063/1.458452}}.

\bibitem{dmol3}
B.~Delley, From molecules to solids with the dmol3 approach, The Journal of
  Chemical Physics 113~(18) (2000) 7756--7764.
\newblock \href {http://dx.doi.org/10.1063/1.1316015}
  {\path{doi:10.1063/1.1316015}}.

\bibitem{Pwc}
J.~P. Perdew, Y.~Wang, Accurate and simple analytic representation of the
  electron-gas correlation energy, Phys. Rev. B 45 (1992) 13244--13249.
\newblock \href {http://dx.doi.org/10.1103/PhysRevB.45.13244}
  {\path{doi:10.1103/PhysRevB.45.13244}}.

\bibitem{PBE1996}
J.~P. Perdew, K.~Burke, M.~Ernzerhof, Generalized gradient approximation made
  simple, Phys. Rev. Lett. 77 (1996) 3865--3868.
\newblock \href {http://dx.doi.org/10.1103/PhysRevLett.77.3865}
  {\path{doi:10.1103/PhysRevLett.77.3865}}.

\bibitem{dspp}
B.~Delley, Hardness conserving semilocal pseudopotentials, Phys. Rev. B 66
  (2002) 155125.
\newblock \href {http://dx.doi.org/10.1103/PhysRevB.66.155125}
  {\path{doi:10.1103/PhysRevB.66.155125}}.

\bibitem{cosmo}
A.~Klamt, G.~Schüürmann, Cosmo: a new approach to dielectric screening in
  solvents with explicit expressions for the screening energy and its gradient,
  J. Chem. Soc.{,} Perkin Trans. 2 (1993) 799--805\href
  {http://dx.doi.org/10.1039/P29930000799} {\path{doi:10.1039/P29930000799}}.

\bibitem{Delleytddft}
B.~Delley, Time dependent density functional theory with {DMol}3, Journal of
  Physics: Condensed Matter 22~(38) (2010) 384208.
\newblock \href {http://dx.doi.org/10.1088/0953-8984/22/38/384208}
  {\path{doi:10.1088/0953-8984/22/38/384208}}.

\bibitem{TDDFT}
E.~Runge, E.~K.~U. Gross, Density-functional theory for time-dependent systems,
  Phys. Rev. Lett. 52 (1984) 997--1000.
\newblock \href {http://dx.doi.org/10.1103/PhysRevLett.52.997}
  {\path{doi:10.1103/PhysRevLett.52.997}}.

\bibitem{mopac}
T.~Hirano, Mopac manual, in: MOPAC Manual, 1993.

\bibitem{wilsonarticle}
W.~Miranda, S.~Coutinho, M.~Tavares, E.~Moreira, D.~Azevedo, Ab initio
  vibrational and thermodynamic properties of adamantane, sila-adamantane
  (si10h16), and c9si1h16 isomers, Journal of Molecular Structure 1122 (2016)
  299 -- 308.
\newblock \href
  {http://dx.doi.org/https://doi.org/10.1016/j.molstruc.2016.05.103}
  {\path{doi:https://doi.org/10.1016/j.molstruc.2016.05.103}}.

\bibitem{Siesta}
J.~M. Soler, E.~Artacho, J.~D. Gale, A.~Garc{\'{\i}}a, J.~Junquera,
  P.~Ordej{\'{o}}n, D.~S{\'{a}}nchez-Portal, The {SIESTA} method forab
  initioorder-nmaterials simulation, Journal of Physics: Condensed Matter
  14~(11) (2002) 2745--2779.
\newblock \href {http://dx.doi.org/10.1088/0953-8984/14/11/302}
  {\path{doi:10.1088/0953-8984/14/11/302}}.

\bibitem{Ceperley1980}
D.~M. Ceperley, B.~J. Alder, Ground state of the electron gas by a stochastic
  method, Phys. Rev. Lett. 45 (1980) 566--569.
\newblock \href {http://dx.doi.org/10.1103/PhysRevLett.45.566}
  {\path{doi:10.1103/PhysRevLett.45.566}}.

\bibitem{borazine}
R.~Boese, A.~H. Maulitz, P.~Stellberg, Solid-state borazine: Does it deserve to
  be entiteled “inorganic benzene”?, Chemische Berichte 127~(10) (1994)
  1887--1889.
\newblock \href {http://dx.doi.org/10.1002/cber.19941271011}
  {\path{doi:10.1002/cber.19941271011}}.

\bibitem{h-bn}
J.~Wang, F.~Ma, M.~Sun, Graphene{,} hexagonal boron nitride{,} and their
  heterostructures: properties and applications, RSC Adv. 7 (2017)
  16801--16822.
\newblock \href {http://dx.doi.org/10.1039/C7RA00260B}
  {\path{doi:10.1039/C7RA00260B}}.

\bibitem{chen2015nanotubes}
Y.~I. Chen, Nanotubes and nanosheets: functionalization and applications of
  boron nitride and other nanomaterials, CRC Press, 2015, p. 443.

\bibitem{KOOPMANS1934104}
T.~Koopmans, Über die zuordnung von wellenfunktionen und eigenwerten zu den
  einzelnen elektronen eines atoms, Physica 1~(1) (1934) 104 -- 113.
\newblock \href
  {http://dx.doi.org/https://doi.org/10.1016/S0031-8914(34)90011-2}
  {\path{doi:https://doi.org/10.1016/S0031-8914(34)90011-2}}.

\bibitem{hardness1}
P.~K. Chattaraj, A.~Poddar, Molecular reactivity in the ground and excited
  electronic states through density-dependent local and global reactivity
  parameters, J. Phys. Chem. A 103 (1999) 8691--8699.
\newblock \href {http://dx.doi.org/https://doi.org/10.1021/jp991214+}
  {\path{doi:https://doi.org/10.1021/jp991214+}}.

\bibitem{parr}
J.-L. Calais, Density-functional theory of atoms and molecules. r.g. parr and
  w. yang, oxford university press, new york, oxford, 1989. ix + 333 pp. price
  £45.00, International Journal of Quantum Chemistry 47~(1) (1993) 101--101.
\newblock \href {http://dx.doi.org/10.1002/qua.560470107}
  {\path{doi:10.1002/qua.560470107}}.

\bibitem{hardness2}
R.~G. Pearson, Absolute electronegativity and hardness: application to
  inorganic chemistry, Inorganic Chemistry 27~(4) (1988) 734--740.
\newblock \href {http://dx.doi.org/10.1021/ic00277a030}
  {\path{doi:10.1021/ic00277a030}}.

\bibitem{hardness}
M.~Hoque, M.~Uzzaman, Physiochemical, molecular docking, and pharmacokinetic
  studies of naproxen and its modified derivatives based on dft, International
  Journal of Scientific Research and Management 6 (2018) 12--19.
\newblock \href {http://dx.doi.org/10.18535/ijsrm/v6i9.c01}
  {\path{doi:10.18535/ijsrm/v6i9.c01}}.

\bibitem{bngap}
J.~H. Kim, T.~V. Pham, J.~H. Hwang, C.~S. Kim, M.~J. Kim, Boron nitride
  nanotubes: synthesis and applications, Nano Convergence 5~(1) (2018) 17.
\newblock \href {http://dx.doi.org/10.1186/s40580-018-0149-y}
  {\path{doi:10.1186/s40580-018-0149-y}}.

\bibitem{bngap2}
X.~Blase, A.~Rubio, S.~G. Louie, M.~L. Cohen, Stability and band gap constancy
  of boron nitride nanotubes, Europhysics Letters ({EPL}) 28~(5) (1994)
  335--340.
\newblock \href {http://dx.doi.org/10.1209/0295-5075/28/5/007}
  {\path{doi:10.1209/0295-5075/28/5/007}}.

\bibitem{BN-optical2}
R.~Arenal, O.~St\'ephan, M.~Kociak, D.~Taverna, A.~Loiseau, C.~Colliex,
  Electron energy loss spectroscopy measurement of the optical gaps on
  individual boron nitride single-walled and multiwalled nanotubes, Phys. Rev.
  Lett. 95 (2005) 127601.
\newblock \href {http://dx.doi.org/10.1103/PhysRevLett.95.127601}
  {\path{doi:10.1103/PhysRevLett.95.127601}}.

\bibitem{Wilson1980}
E.~Wilson, J.~Decius, P.~Cross, Molecular Vibrations: The Theory of Infrared
  and Raman Vibrational Spectra, Dover Books on Chemistry Series, Dover
  Publications, 1980.

\bibitem{Porezag1996}
D.~Porezag, M.~R. Pederson, Infrared intensities and raman-scattering
  activities within density-functional theory, Phys. Rev. B 54 (1996)
  7830--7836.
\newblock \href {http://dx.doi.org/10.1103/PhysRevB.54.7830}
  {\path{doi:10.1103/PhysRevB.54.7830}}.

\bibitem{Moreira}
E.~Moreira, J.~Henriques, D.~Azevedo, E.~Caetano, V.~Freire, E.~Albuquerque,
  Structural, optoelectronic, infrared and raman spectra of orthorhombic srsno3
  from dft calculations, Journal of Solid State Chemistry 184~(4) (2011) 921 --
  928.
\newblock \href {http://dx.doi.org/https://doi.org/10.1016/j.jssc.2011.02.009}
  {\path{doi:https://doi.org/10.1016/j.jssc.2011.02.009}}.

\bibitem{Moreira1}
E.~Moreira, J.~M. Henriques, D.~L. Azevedo, E.~W.~S. Caetano, V.~N. Freire,
  U.~L. Fulco, E.~L. Albuquerque, Structural and optoelectronic properties, and
  infrared spectrum of cubic basno3 from first principles calculations, Journal
  of Applied Physics 112~(4) (2012) 043703.
\newblock \href {http://dx.doi.org/10.1063/1.4745873}
  {\path{doi:10.1063/1.4745873}}.

\bibitem{jefs2013}
J.~E. F.~S. Rodrigues, E.~Moreira, D.~M. Bezerra, A.~P. Maciel, C.~W. [de
  Araujo~Paschoal], Ordering and phonons in ba3canb2o9 complex perovskite,
  Materials Research Bulletin 48~(9) (2013) 3298 -- 3303.
\newblock \href
  {http://dx.doi.org/https://doi.org/10.1016/j.materresbull.2013.05.005}
  {\path{doi:https://doi.org/10.1016/j.materresbull.2013.05.005}}.

\bibitem{Moreira2}
J.~Henriques, C.~Barboza, E.~Albuquerque, U.~Fulco, E.~Moreira, Structural,
  optoelectronic, infrared and raman spectra from first principles calculations
  of $\gamma$-cd(oh)2, Journal of Physics and Chemistry of Solids 76 (2015) 45 -- 50.
\newblock \href {http://dx.doi.org/https://doi.org/10.1016/j.jpcs.2014.08.003}
  {\path{doi:https://doi.org/10.1016/j.jpcs.2014.08.003}}.

\bibitem{Moreira3}
E.~Moreira, C.~Barboza, E.~Albuquerque, U.~Fulco, J.~Henriques, A.~Araújo,
  Vibrational and thermodynamic properties of orthorhombic casno3 from dft and
  dfpt calculations, Journal of Physics and Chemistry of Solids 77 (2015) 85 --
  91.
\newblock \href {http://dx.doi.org/https://doi.org/10.1016/j.jpcs.2014.09.016}
  {\path{doi:https://doi.org/10.1016/j.jpcs.2014.09.016}}.

\bibitem{Zhukovskii}
Y.~F. Zhukovskii, S.~Piskunov, N.~Pugno, B.~Berzina, L.~Trinkler, S.~Bellucci,
  Ab initio simulations on the atomic and electronic structure of single-walled
  bn nanotubes and nanoarches, Journal of Physics and Chemistry of Solids
  70~(5) (2009) 796 -- 803.
\newblock \href {http://dx.doi.org/https://doi.org/10.1016/j.jpcs.2009.03.016}
  {\path{doi:https://doi.org/10.1016/j.jpcs.2009.03.016}}.

\bibitem{Hamdi}
I.~Hamdi, N.~Meskini, Ab initio study of the structural, elastic, vibrational
  and thermodynamic properties of the hexagonal boron nitride: Performance of
  lda and gga, Physica B: Condensed Matter 405~(13) (2010) 2785 -- 2794.
\newblock \href {http://dx.doi.org/https://doi.org/10.1016/j.physb.2010.03.070}
  {\path{doi:https://doi.org/10.1016/j.physb.2010.03.070}}.

\bibitem{Erba}
A.~Erba, M.~Ferrabone, J.~Baima, R.~Orlando, M.~Rérat, R.~Dovesi, The
  vibration properties of the (n,0) boron nitride nanotubes from ab initio
  quantum chemical simulations, The Journal of Chemical Physics 138~(5) (2013)
  054906.
\newblock \href {http://dx.doi.org/10.1063/1.4788831}
  {\path{doi:10.1063/1.4788831}}.

\bibitem{Wirtz}
L.~{Wirtz}, A.~{Rubio}, Vibrational properties of boron-nitride nanotubes:
  effects of finite length and bundling, IEEE Transactions on Nanotechnology
  2~(4) (2003) 341--348.
\newblock \href {http://dx.doi.org/10.1109/TNANO.2003.820511}
  {\path{doi:10.1109/TNANO.2003.820511}}.

\bibitem{Hasi}
F.~Hasi, F.~Simon, H.~Kuzmany, R.~Arenal de~la Concha, A.~Loiseau, Raman
  spectroscopy of boron nitride nanotubes and boron nitride — carbon
  composites, AIP Conference Proceedings 786~(1) (2005) 340--344.
\newblock \href {http://dx.doi.org/10.1063/1.2103883}
  {\path{doi:10.1063/1.2103883}}.

\bibitem{Arenal}
R.~Arenal, A.~C. Ferrari, S.~Reich, L.~Wirtz, J.-Y. Mevellec, S.~Lefrant,
  A.~Rubio, A.~Loiseau, Raman spectroscopy of single-wall boron nitride
  nanotubes, Nano letters 6~(8) (2006) 1812--1816.
\newblock \href {http://dx.doi.org/https://doi.org/10.1021/nl0602544}
  {\path{doi:https://doi.org/10.1021/nl0602544}}.

\bibitem{Nemanich}
R.~J. Nemanich, S.~A. Solin, R.~M. Martin, Light scattering study of boron
  nitride microcrystals, Phys. Rev. B 23 (1981) 6348--6356.
\newblock \href {http://dx.doi.org/10.1103/PhysRevB.23.6348}
  {\path{doi:10.1103/PhysRevB.23.6348}}.

\bibitem{BN-optical}
J.~S. Lauret, R.~Arenal, F.~Ducastelle, A.~Loiseau, M.~Cau, B.~Attal-Tretout,
  E.~Rosencher, L.~Goux-Capes, Optical transitions in single-wall boron nitride
  nanotubes, Phys. Rev. Lett. 94 (2005) 037405.
\newblock \href {http://dx.doi.org/10.1103/PhysRevLett.94.037405}
  {\path{doi:10.1103/PhysRevLett.94.037405}}.

\bibitem{kim2018boron}
J.~H. Kim, T.~V. Pham, J.~H. Hwang, C.~S. Kim, M.~J. Kim, Boron nitride
  nanotubes: synthesis and applications, Nano convergence 5~(1) (2018) 17.
\newblock \href {http://dx.doi.org/https://doi.org/10.1186/s40580-018-0149-y}
  {\path{doi:https://doi.org/10.1186/s40580-018-0149-y}}.

\end{thebibliography}
\end{document}